# Dye-Encapsulated Zeolitic Imidazolate Framework (ZIF-71) for Fluorochromic Sensing of Pressure, Temperature, and Volatile Solvents

*Yang Zhang, Mario Gutiérrez, Abhijeet K. Chaudhari, and Jin-Chong Tan**

Y. Zhang, Dr. M. Gutiérrez, Dr. A.K. Chaudhari, Prof. J.C. Tan
Multifunctional Materials & Composites (MMC) Laboratory, Department of Engineering Science, University of Oxford, Parks Road, Oxford OX1 3PJ, United Kingdom.

*Corresponding author
E-mail: jin-chong.tan@eng.ox.ac.uk

**Keywords:** Hybrid materials; metal-organic frameworks; luminescence; guest-host interactions; photonic sensors

**Abstract:**

Luminescent metal-organic frameworks (MOFs) offer a multifunctional platform for creating non-invasive sensors and tuneable optoelectronics. However, fluorochromic materials that are photophysically resilient and show high sensitivity towards different physical and chemical stimuli are scarce. We report a facile host-guest nanoconfinement strategy to construct a fluorescent hybrid material with multiple sensing capabilities. We design and fabricate a new Guest@MOF material: comprising a zeolitic MOF (ZIF-71) as a nanoporous host for encapsulating rhodamine B (RhB dye) guest molecules, resulting in the RhB@ZIF-71 system with mechanochromic, thermochromic, and solvatochromic sensing response. The fluorochromic sensing properties stem from the nanoconfinement effect that ZIF-71 imposes on RhB monomers, yielding the H- or J-type aggregates with tuneable photophysical and photochemical properties. For mechanochromism, the external pressure causes an emission red shift in a linear fashion, switching RhB guests from H-type to J-type aggregates through a shear deformation. For thermochromism, we demonstrate a linear scaling as a function of temperature due to the spatial restriction imposed on J-type aggregates incarcerated in ZIF-71 pores. Harnessing the solvatochromism of RhB@ZIF-71, we identified three diverse groups of volatile organic compounds. The multimodal sensing response pave the way to smart applications like photonic pressure sensors, non-invasive thermometers, and ultrasensitive chemosensors.

## 1. Introduction

Research on the luminescent sensing abilities of metal-organic framework (MOF) materials and their composite systems is rapidly expanding, because of its wide range of technological applications in nanophotonics, bio-imaging, and smart sensors.[1-4] Compared to other luminescent sensing materials such as metal complexes[5] and dye-based fluorescent probes,[6] crystalline MOF exhibits long-range periodicity, nanosized pores, combined with diverse chemical and structural versatility for tuning its vast physicochemical properties.[7] Among the many reported luminescent MOFs,[8] the study of Guest@MOF systems,[9-14] composed of a MOF structure (the "host") to afford the encapsulation of luminescent molecules (the "guest"), has attracted considerable attention. This is because the Guest@MOF composite can be achieved by an easy-to-synthesise or 'off-the-shelf' MOF structure, and commonly available fluorophores (luminescent dyes) to yield bespoke luminescent properties.[15-17] Additionally, the MOF hosts will act as a shield, protecting the guest against possible sources of degradation, like in the case of the encapsulation of the unstable fluorescent perovskite nanocrystals, which enhanced their robustness when confined within MOFs, paving the way to their integration into functional photonic devices.[18-20] Notably, the aforementioned approach could circumvent the need to rely on complex chemical designs to yield luminescent MOFs, or to incorporate expensive rare-earth elements widely deployed in inorganic phosphors.[21]

The existing Guest@MOF approach faces a number of challenges. For example, the vast majority of Guest@MOF sensing mechanisms are based on the luminescence quenching effect, where the analytes 'switch off' the emission of the guest.[22-24] However, this mechanism might be susceptible to moisture exposure or temperature fluctuation in the environment giving inaccurate readouts.[15, 25] There are other Guest@MOF materials exhibiting luminescent sensing through a 'turn-on' sensing[26] or luminescence color (fluorochromic) switching mechanism, but they often require very specific chemical interactions to function[15] thus limiting multimodal applications. To address some of the foregoing challenges, we propose that fluorochromic sensing can be achieved by using MOF pores to induce different caging effects on luminescent dye monomers and their aggregates, thereby producing distinctive fluorescent behavior when subject to different physical and chemical stimuli.

In this study, we demonstrate a facile method to synthesize a multi-stimuli responsive Guest@MOF luminescent material, by embedding the well-known fluorescent dye rhodamine B (RhB) into zeolitic imidazolate framework-71 (ZIF-71). Although a number of RhB@MOF systems have been reported,[27-34] there are certain limitations, such as the need to use complex synthesis steps, no control of RhB aggregates, inadequate understanding of the underlying mechanisms, and single-mode sensing functionality (basically only solvatochromism). In contrast, our RhB@ZIF-71 not only shows solvatochromic response with its sensitivity far surpassing all reported examples,[27-34] but also exhibits intriguing mechanochromic and thermochromic sensing properties that are unknown to date. We employed detailed photophysical characterization techniques to unravel the underpinning guest-host mechanisms controlling the performance of this new composite system. Our results also demonstrate proof-of-concept RhB@ZIF-71 applications opening new pathways for the realization of multimodal fluorescent sensing platforms.

## 2. Result and Discussion

### 2.1 Synthesis and Structure of the RhB@ZIF-71 System

The RhB@ZIF-71 system was synthesized through rapid mixing of a solution of zinc acetate with a solution of 4,5-dichloroimidazole (dcIm) and RhB at room temperature; the full details are given in the Experimental Section. Through this straightforward one-step reaction, we observe that the mixed solution was immediately converted from transparent pink color to turbid, indicating the rapid formation of RhB@ZIF-71. As shown in Figure 1, the crystal size of RhB@ZIF-71 was found to be ~800 nm by scanning electron microscopy (SEM) and atomic force microscopy (AFM). We also characterized the morphology of RhB@ZIF-71 at different reaction times (Figure S1, Supporting

Information), and found after ~10 minutes, a crystal size of ~800 nm has formed. The rapid formation of such sub-micron sized crystals is consistent with the previous reports on pure ZIF-71.[35]

Powder X-ray diffraction (PXRD) was carried out to verify the crystal structure. As shown in Figure 2a and Figure S2, the PXRD patterns of RhB@ZIF-71 with different RhB concentrations and pure ZIF-71 were consistent with the simulated XRD pattern of ZIF-71, indicating that the introduced guest (RhB) does not affect the structure of ZIF-71 significantly. However, when a very high concentration of RhB (0.5 mmol) was used, we observed the appearance of the (001) peak at $2\theta = 3.1°$ (Figure 2a inset), while this peak was not obvious in the samples whose RhB concentrations are relatively low (0.01 and 0.05 mmol). Although the peak at $2\theta = 3.1°$ exists in the simulated pattern, its relative intensity is typically too low to be observed by XRD. Since a smaller $2\theta$ value represents a larger crystal plane separation, the appearance of the (001) peak suggests that many ZIF-71 pores may contain more than one RhB molecules, thereby affecting the preferred orientation during crystal growth. Because the size of one RhB molecule is around 16.19 × 12.83 × 6.97 Å (including van der Waals surface) and the minimum/maximum distance inside the ZIF-71 pore is 16.58 / 22.59 Å, it is conceivable that the crystal growth preference will be influenced when more than one RhB molecules are stacked together. This is further supported by the results of thermogravimetric analysis (TGA) (Figure 2b), from which the determined chemical formula shows there was on average more than one RhB per pore.

Although we did not see the trace of RhB in the PXRD patterns (Figure S2), it was successfully detected by Fourier-transform infrared spectroscopy with attenuated total reflection (ATR-FTIR) (Figure 2c), TGA (Figure 2b), and Raman vibrational spectroscopy (Figure S3). Figure 2b and 2c demonstrate the concentration of RhB inside the RhB@ZIF-71 system gradually increases as the amount of RhB used in the synthesis was increased, as evidenced in the systematic rise of the ~1590 $cm^{-1}$ mode shown in Figure 2c inset (left). Intriguingly, from Figure 2c inset (right), we established that the peak at ~1710 $cm^{-1}$ attributed to the C=O vibrational mode of RhB has completely disappeared after confinement. Likewise, we detected the similar phenomena in the Raman spectra (Figure S3). Vibrational spectroscopic data revealed that the C=O of RhB could interact with the zinc atoms of ZIF-71 or with its dcIm linkers. The TGA results (Figure 2b) not only enabled us to derive the chemical formula of RhB@ZIF-71, but also revealed the enhanced thermal stability of RhB when confined in the system; this finding supports the notion that RhB guests are residing in the pores of ZIF-71.

**2.2 Luminescent Properties of the RhB@ZIF-71 System and its Constituents**

We begin by investigating the emission properties of the pristine crystals of ZIF-71 and its dcIm linker under room temperature, as depicted in Figures 3a and 3b. It can be seen in Figure S4 that, the dcIm linker displays an intense and broadband emission in the solid-state state with emission maximum at ~468 nm (under 360 nm UV excitation), which can be attributed to the $\pi^* - \pi$ transition.[36] From the emission map (Figure 3c) and the emission spectra (Figure S4) of ZIF-71, we established that ZIF-71 exhibits two emission peaks at around 456 nm and 559 nm. Because neither dcIm nor Zn(II) has emission at 559 nm, the 559 nm emission of ZIF-71 could be from the ligand-metal charge transfer (LMCT), and the 456 nm emission comes from the linker itself in ZIF-71. Previous literature[37] mentioned that LMCT process is usually expressed in MOF containing Zn(II), especially when the linker contains benzene derivatives, and MOFs often emit green color fluorescence (500 - 565 nm) when LMCT occurs. Clearly, the structure and performance of ZIF-71 almost completely conform to this commonality, which supports our reasoning.

To further confirm that, we measured the emission lifetimes of dcIm (Table 1) and ZIF-71 (Table 2) employing the time-correlated single-photon-counting (TCSPC) technique. The most noticeable variation is when the observed wavelength changed from 450 nm to 558 nm (in Table 2), the $c_3$ of ZIF-71 dramatically increased, which we attribute to the effect of LMCT. However, it is not certain that $\tau_3$ is the lifetime of LMCT, because the $\tau_3$ of dcIm itself is 4.44 ns (in Table 1) and after forming the ZIF-71 framework structure, theoretically, it will increase due to caging effect.[38] The increase in $\tau_3$ of dcIm itself may be very close to the lifetime of the LMCT, which may cause the two to become indistinguishable. Thus, we consider the $\tau_3$ of dcIm itself and the lifetime of LMCT together constitute the $\tau_3$ of ZIF-71.

Subsequently, we characterized the band gap (Figure 3d), solid-state excitation and emission spectra (Figures 3e and 3f), absorption (Figure S5) of RhB@ZIF-71, as well as the emission spectra (Figure S6) of pure RhB solution with different concentrations at room temperature. However, the emission of all the RhB@ZIF-71 powders is dominated by the guest itself (rather than simultaneously manifested by emissions of the guest and the LMCT of ZIF-71). The other possible interactions, involving C=O with the zinc atoms or the linkers, or the possible interaction between nitrogen atoms/the xanthene ring of RhB and the open metal sites/ linker,[39] may interrupt the LMCT process causing the single emission peak of RhB@ZIF-71. Of course, we do not rule out the possibility that emission of ZIF-71 host could not be observed due to its relatively low quantum yield (Table S1).

In terms of their emission spectra (Figure 3f), it is shown that the bathochromic (red) shift was detected when the concentration of RhB increased. We propose that one of the reasons for this red shift is related to the formation of more RhB aggregates. Because the relative dimensions of RhB molecules and ZIF-71 pores, the XRD results (Figure 2a), and the chemical composition derived from TGA results (Figure 2b) suggest that more than one RhB molecules may occupy the pore of ZIF-71. It follows that more RhB introduced during synthesis will lead to more aggregates. In principle, both of H-type (head-to-head) and J-type (head-to-tail) aggregates of RhB are able to form during the synthesis.[40, 41] Since the emission of H-type aggregates is theoretically forbidden and the emission of J-type aggregates is allowed but with a longer wavelength,[42] we could observe the red shift. This idea is confirmed by the excitation spectra in Figure 3e. It can be seen that there are three excitation areas: humps at around 495 nm; peaks at 531 nm; and shoulders located between 552 - 590 nm. Using Kasha's exciton model,[42] we assigned the humps to H-type aggregates, peaks to RhB monomers, and the shoulders to J-type aggregates. Compared with the peaks (~531 nm), the intensity of the humps (~495 nm) increased so it follows that the amount of aggregates increased when a higher concentration of RhB was used during synthesis. The reason why the shoulders (552 - 590 nm) did not increase is explained below.

Comparing the variation of H-type and J-type aggregates, we found that, the excitation peaks of the J-type aggregates showed a red shift as the concentration of RhB increases, while the H-type did not show a blue shift. This difference arises because the J-type aggregates have a longer spatial dimension than the H-type aggregates,[41] which might allow the J-type aggregates to interact with another J-type aggregate in adjacent pores. Our group has reported that perylene@ZIF-8 also showed a similar phenomenon, in which perylene and 2-methylimidazole can form an energy transfer pathway through the adjacent pores.[43] Likewise, other researchers have demonstrated the preparation of long-range crystalline MOFs by mechanochemistry under solid conditions,[44, 45] which means that chemical reactions can occur between solid crystals. On this basis, weak interactions involving J-aggregate interaction across the pores is plausible. Moreover, because the maximum size of ZIF-71 window aperture is 5.08 Å, which is spatially larger than some parts of the RhB molecules (e.g. the distance of C-C on the xanthene ring is 4.79 Å), we suggest that part of the J-type aggregates may protrude out of the ZIF-71 window, which will strengthen the interaction by bridging the pores. In contrast, the packing of H-type aggregates is tighter, and the occupied space is relatively small,[41] allowing them to be better confined inside the pores and less likely to interact with guests in adjacent pores. Hence, when the concentration of RhB used in the synthesis was increased, it will cause more J-type aggregates-based interactions (bridging the pores) in the RhB@ZIF-71 system, which leads to a smaller band gap (Figure 3d). This is another reason for the observed red shift in emission, and the reason behind the declining intensity of the emission shoulders (552 - 590 nm in Figure 3e).

To further study the formation of aggregates and analyse their luminescent properties, the emission lifetimes of pure RhB and RhB@ZIF-71 were measured by TCSPC technique. For pure RhB in MeOH solution (0.0001 M), the lifetime is 2.93 ns and for all the RhB@ZIF-71 powders, we obtained three decay times as summarized in Figure 3g (see also Table S2 and Figure S7). We propose that $\tau_1$ can be assigned to the H-type aggregates, $\tau_2$ corresponds to the J-type aggregates, and $\tau_3$ is due to the RhB monomers. Firstly, it can be seen that $\tau_3$ is greater than 2.93 ns, which also is evidence that RhB is residing inside the pore, because the vibration of RhB monomers confined in the pore becomes restricted (i.e. caging effect) greatly reducing their non-radiative decay, and thus increasing the lifetime. Secondly, it can also be seen that as the RhB concentration increases, the values of $a_1$ and $c_1$ rise while $a_3$ and $c_3$ fell (Table S2), which indicates an increase in the content of H-type aggregates and a relative decrease in monomer content. Thirdly, $\tau_2$ was found to decrease as the RhB concentration increases, this supports the notion that the J-type aggregates could interact with each other across the adjacent pores. This kind of interaction can also be proven by comparing $a_2$ and $c_2$ (Table S2). As the observed wavelength ($\lambda_{\text{obs}}$) increases, the magnitude of the change in $a_2$ and $c_2$ decreases, which results in more interactions and hence broadening of the emission component of the J-type aggregation. However, comparing the changes in $a_1$, $c_1$, $a_2$ and $c_2$, we can see that the increase of J-type aggregates is not as high as the H-type aggregates. Given the space limitation of the ZIF-71 pore, the larger J-type aggregates are harder to form than the H-type. Herein, the lifetime data have substantiated the previous inferences obtained from the emission and excitation spectrum.

The quantum yield of $RhB_I$@ZIF-71, $RhB_{II}$@ZIF-71, $RhB_{III}$@ZIF-71 were characterized, and the results are summarized in Table 3.

### 2.3 Mechanochromic Sensing Response

To study the mechanochromism of the RhB@ZIF-71 system, the $RhB_{II}$@ZIF-71 material was chosen and compressed into pellets under different pressures. Figures 4a and 4b depict the color of the pellets viewed under day light and their emissions when subject to a 365-nm UV excitation, respectively. Here we focus on the $RhB_{II}$@ZIF-71 pellets, because the $RhB_{III}$@ZIF-71 has a relatively low quantum yield (Table 3), while the emission of the $RhB_I$@ZIF-71 pellets (Figure S8) is not as linear as $RhB_{II}$@ZIF-71. Figure 4d and Figure S9b reveal that the emission spectra of these pellets red-shifted as the pressure was systematically raised up to ~350 MPa, demonstrating a very linear relationship that is highly desirable for stress sensing applications (Figure 4f).

Combined with the excitation spectra (Figures 4c and S9a), the lifetime data (Figures 4g and S10, Table S3), and the PXRD patterns (Figures 4e and S11) of the pellets, we investigate the reason for the observed red shift in emission. On the one hand, it can be seen that the excitation peaks of J-type aggregates (575-590 nm) showed a red shift with increasing pressure, and in the PXRD pattern (Figure 4e and S11) it can be seen that the pressure has caused some amorphization of the ZIF-71 structure. These results suggest that the pelleting pressure leads to the mechanical deformation of the ZIF-71 structure, where framework distortion will cause the tighter packing of aggregates and make the adjacent pores to come closer, causing stronger interactions inside the pores, and promoting stronger interactions between the J-type aggregates across adjacent pores. Together, these factors result in a red shift. In principle, the stronger the interaction, the shorter the luminescent lifetime becomes,[46] but the $\tau_2$ in Figure 4g did not decrease. This is because RhB aggregates are present in the pores of ZIF-71, hence the pore shrinkage from mechanical stress introduces a stronger caging effect, suppressing the non-radiative decay and preventing the decrease of lifetime. Based on this hypothesis, it is easy to understand the increase of the monomer's lifetime inside the pores ($\tau_3$), which is dependent only upon the caging effect.

On the other hand, we consider the red shift of the emission is also related to the increase in the relative content of the J-type aggregates. Table S4 shows that the RhB@ZIF-71 pellets had a smaller FWHM (full width at half maximum) than the ZIF-71 pellets under pressure, which means that the RhB@ZIF-71 pellets possess a higher crystallinity than ZIF-71 (Figure S12) and reveals that the encapsulated guests can mechanically enhance the structural stability of the overall framework under stress. Therefore, we propose the reason why $a_3$ and $c_3$ in Table S3 started to drop significantly at relatively low pressure, $a_1$ $c_1$ rose first and then fell, and $a_2$ $c_2$ continuously increased is that the pores containing monomers are initially destroyed to form new H-type and J-type aggregates at relatively low pressure. Subsequently, as the pressure keeps rising, the ZIF-71 crystals continue to deform under shear deformation,[47] causing the aggregates to transform from H-type to J-type (Scheme 1). Moreover, in the excitation spectra of the $RhB_{II}$@ZIF-71 pellets (Figure 4c), the intensity of the H-type aggregates first rose and then fell, but the peak position of H-type aggregates (450 - 506 nm) is always blue shifted with increasing pressure, which also supports this hypothesis.

### 2.4 Thermochromic Sensing Response

RhB@ZIF-71 exhibits thermochromic behavior as a function of temperature, here we use $RhB_{II}$@ZIF-71 as an example to explain the underlying mechanism. Figure 5(a-b) shows the excitation and emission spectra from room temperature to 200 °C, respectively. As shown in Figure 5 and Figure S13, the luminescent intensity of RhB@ZIF-71 decreases accompanied by a red shift with increasing temperature. The decrease in intensity is very similar to the performance of pure RhB itself in solution at different temperatures,[48, 49] which is due to the increase in non-radiative decay rate. In the previous research,[48] the temperature range of this kind of intensity decrease of RhB was generally from 5 °C to 80 °C. Above this temperature range, the luminescent intensity of RhB was too low to be detected accurately. Remarkably, we demonstrate that the non-radiative decay of RhB was greatly reduced due to the caging effect of ZIF-71, not only overcoming the restriction that thermochromism of RhB can only be achieved in solutions, but our solid-state system significantly extends the operational temperature range by at least a factor of two (Figure 5d). Figure S13d shows that when the temperature rises from 200 °C to 250 °C, RhB@ZIF-71 experienced a relatively large red shift, which can be attributed to the thermal decomposition of RhB itself beyond 200 °C (consistent with TGA results in Figure 2b).

By analyzing the temperature range from room temperature to 200 °C, we established that there is a linear scaling relationship between the red shift of the emission peak and the temperature increment (Figure 5d); this effect is highly attractive for photonics-based thermometry applications. Note that the red shift of RhB in the solid state observed here is as yet unreported in the literature; unlike for pure RhB solutions, in which the increased in temperature has produced no red shift.[48] The normalised excitation spectra (Figure 5a) reveal that, in this linearly changing region (room temperature to 200 °C), the relative intensity of H-type aggregates and monomers did not change much with the increase of temperature, while the J-type aggregates showed a relatively large intensity enhancement at high temperatures (i.e. 150 °C and 200 °C). Since the crystal structure of RhB@ZIF-71 can withstand a temperature up to 250 °C (evidenced from the XRD patterns in Figure 5c and Figure S13e), we reason that the different spatial sizes between the H- and J-type aggregates, and its monomers determine the trend of their intensity and wavelength change.

As discussed above, H-type aggregates and monomers possess smaller size than J-type aggregates, so the J-type aggregates will experience a stronger caging effect, therefore becoming less sensitive to temperature variation. In other words, when temperature rises, the excitation intensity of J-type aggregates decreases less than that of H-aggregates and monomers. It is this effect that causes the red shift observed in the emission peak. Additionally, at high temperatures, it can be seen that the excitation peaks of the H-type and J-type aggregates showed a very small degree of red and blue shifts, respectively, but the peaks associated with the monomers remain unchanged. This indicates that both H-type and J-type aggregates slightly expand at high temperatures, thereby weakening the interactions within themselves. The absence of any variation to the excitation peaks of monomers reveals that the interactions between RhB and ZIF-71 are weak or negligible. Likewise, in the mechanochromism studies, we note that the peak wavelength of the monomer also did not change subject to mechanical stress (Figure 4c), which also indicate the interactions are weak or negligible.

### 2.5 Solvatochromic Sensing Response

Another promising property of RhB@ZIF-71 is its solvatochromic response as shown in Figure 6 and Figure S14. In order to understand the subtle changes of emission due to solvatochromism, we chose the $RhB_I$@ZIF-71 system which has the lowest RhB concentration. We observed that the peak intensity and peak position of $RhB_I$@ZIF-71 were distinctively different when exposed to different solvents. To explain this phenomenon, we propose that this solvatochromic response is linked to the nature of RhB itself. Because RhB can exist in three different forms in solution state (i.e. lactone, zwitterion, and cation):[48] the lactone has no color under visible light or UV, while the zwitterion and cation have luminescence but their emission wavelength and intensity are different. Generally, polar protic solvents can stabilize the zwitterion,[48] and therefore, will give to an intense luminescence. Whereas, in highly polar aprotic solvents, such as dimethylformamide (DMF), RhB has no luminescence due to the complete conversion to lactone; and in less polar aprotic solvents, such as acetonitrile (ACN), it can show luminescence.[49] Thereby, RhB is solvatochromic,[48, 50] and we propose this is also the basis for the solvatochromism observed in the RhB@ZIF-71 system.

But after confinement within ZIF-71, we found that the luminescence of RhB@ZIF-71 is greatly enhanced compared with the pure RhB (Figure S14). Notably, pure RhB has poor solubility or it is simply insoluble in many non-polar solvents. For example, in this study we tested toluene, hexane, and cyclohexane and confirmed that there was no luminescence in these hydrocarbons due to the aforementioned limitations. Conversely, we discovered that the RhB@ZIF-71 crystals exhibit good dispersion and luminescence (Figure S14 and Figure 6) in hexane and also in cyclic hydrocarbons (e.g. toluene and cyclohexane), thereby demonstrating sensing property previously not achievable by pure RhB alone. To date, most of the RhB@MOF studies[27-34] have focused on the field of solvatochromism, and many of these systems possess the similar sensing behavior. But almost all the authors attributed this kind of sensing to different solvents that affect the energy transfer between the MOF they used and RhB, and lack of in-depth understanding of the emission wavelength change. Moreover, their theory is also unable to satisfactorily explain the luminescence of RhB@MOF systems in some highly polar aprotic solvents. Because no matter how the energy transfer occurs, when RhB turns into the lactone in strong polar aprotic solvents, in principle, the luminescence of the systems will be largely deteriorated.

Here, our systematic analysis of RhB@ZIF-71 based on the new confinement strategy provides new insight into the complex solvatochromism of RhB@MOF systems, as can be seen in Figure 6d. All RhB@ZIF-71 samples in polar aprotic solvents exhibit a longer wavelength accompanied by a reduced intensity than in polar protic solvents. Remarkably, compared with the poor luminescence of RhB in highly polar aprotic solvents, the introduction of ZIF-71 greatly improves the luminescence of RhB (Figure S14). This phenomenon can be explained *via* the concept of aggregates we introduced above. When the pores of ZIF-71 contain monomers, there is enough room for the monomers to contact the aprotic solvent molecules to convert to lactone; when the pores contain aggregates (especially the J-type), it is likely that the remaining space in the pores cannot accommodate the solvents molecules. In other words, the ZIF-71 (host) can protect or shield the aggregates from direct exposure to the solvent molecules. On this basis, compared with the protic solvents, much more monomers inside the ZIF-71 pores converts to the colorless lactone in aprotic solvents, but the luminescence of aggregates is better protected. Consequently, RhB@ZIF-71 in polar aprotic solvents can exhibit luminescence with a longer wavelength and relatively smaller intensity. This is the reason why RhB@ZIF-71 can show a better luminescence and sensing performance than pure RhB alone. Additionally, our interpretation can also be confirmed by observing the luminescent response of RhB@ZIF-71 subject to polar protic solvents (Figure S14c), e.g. methanol (MeOH), ethanol (EtOH), and isopropanol (IPA), because Figure S14c reveals that the emission wavelength of RhB@ZIF-71 changes less than pure RhB especially for IPA, which again can be attributed to the ZIF-71's protection of the RhB aggregates.

## 3. Conclusions

In summary, through a facile guest-host nanoconfinement strategy performed at ambient conditions, we demonstrate the encapsulation of fluorescent RhB monomers (or switchable aggregates) caged within the pores of ZIF-71. The new RhB@ZIF-71 system not only allows RhB to easily yield luminescence in the solid state, but also provides remarkable mechanochromism, thermochromism, and solvatochromism properties that are not achievable to date by traditional use of RhB dispersion in the liquid state, or, indeed by any other means of RhB@MOF systems known thus far. Above all, in the process of analyzing mechanochromism, thermochromism, and solvatochromism, we found several unique mechanisms summarized below. (i) Under mechanical stress, the RhB@ZIF-71 crystals deform by shear causing the conversion of H-type aggregates into J-type aggregates inside the MOF pores, giving rise to pressure sensing. (ii) The J-type aggregates are less affected by temperature due to the strong caging effect provided by the ZIF-71 pores, resulting in the red shift of emission in a very linear fashion, giving rise to non-invasive temperature sensing. (iii) The protective effect of ZIF-71 pores reduces the influence of the solvents has on the RhB aggregates, leading to solvatochromic sensing of volatile organic compounds — previously undetected by unconfined RhB dyes alone.

Our RhB@ZIF-71 nanoconfinement strategy demonstrates the concept of a Guest@MOF system with multimodal fluorescent sensing response, presenting a new platform for the design of smart luminescent sensors. Moreover, we show that the exploitation of luminescent aggregates confined in MOF pores, studied through the characterization of excitation spectrum combined with the luminescent lifetime data is a powerful approach for understanding the mechanisms of novel fluorochromic materials like the dye-encapsulated composite exemplified in this study.

## 4. Experimental Section

### 4.1 Synthesis of RhB@ZIF-71

90 mL methanol clear solution of 2.40 mmol zinc acetate was rapidly poured into 90 mL methanol solution of 9.60 mmol dcIm and different amount of RhB (I: 0.01 mmol; II: 0.05 mmol; III: 0.5 mmol) under stirring. The mixed solution immediately changed from clear to turbidity. After 24 hours of stirring at room temperature, the sample was centrifuged at 8000 rpm to remove the excess reactants, and subsequently washed twice with methanol (the sample was first put into methanol, followed by sonication for 10 minutes, and then the product was separated by centrifugation) to remove the excess reactants and any RhB adhered on the surface of ZIF-71. We observed that, for the $RhB_I$@ZIF-71 (0.01 mmol) and $RhB_{II}$@ZIF-71 (0.05 mmol) samples, the methanol became clear after the second washing cycle, indicating the negligible loss of RhB guest molecules from the ZIF-71 host. Conversely, the $RhB_{III}$@ZIF-71 (0.5 mmol) sample has RhB molecules that remained on the surface after two or more washing cycles, as evidenced from the pink colour of the methanol solvent after multiple washes. The latter indicates surface adhesion of RhB molecules when a high concentration of guest was being introduced during the synthesis step.

The procedure for preparing ZIF-71 was the same, except no RhB was added during the synthesis process.

### 4.2 Sample Preparation for Fluorochromic Characterization:

1) Mechanochromism: the powders were pressed into pellets by using a manual hydraulic pressure equipment with a 1 cm diameter die under a force of 1 ton, 2 ton, 3 ton, and 4 ton. 2) Thermochromism: No extra preparation was required. 3) Solvatochromism: the concentration of 1 mg RhB@ZIF-71 in 20 mL solvents was chosen to avoid the negative influence of too intense emission. Then in order to compare the performance of pure RhB and RhB@ZIF-71 in different solvents, several different concentrations of RhB-MeOH solutions were tested, and a concentration of $7.5 \times 10^{-6}$ M was selected, at which the emission peak wavelength of pure RhB was identical to that of 1 mg RhB@ZIF-71 in 20 mL MeOH. Subsequently, the solutions of pure RhB in different solvents were also formulated at this concentration.

**4.3 Materials Characterization**

The structures and morphologies were examined under scanning electron microscopy (SEM, Carl Zeiss EVO LS15) and atomic force microscopy (AFM, Veeco Dimension 3100). FS5 spectrofluorometer (Edinburgh Instruments) was used to characterize the steady-state emission, excitation spectra, QY, CIE 1931 and lifetime measurements (more details described in the SI). FTIR and Raman results were recorded by using Nicolet iS10 FTIR spectrometer, and MultiRam FT-Raman spectrometer (Bruker), respectively. UV-2600 UV-vis spectrophotometer (Shimadzu) was used to measure the absorption spectra and calculate the Kubelka-Munk (K-M) function. PXRD pattern was recorded using the Rigaku MiniFlex with a Cu Kα source (1.541 Å). TGA was performed using the TGA-Q50 machine (TA Instruments) equipped with a platinum sample holder under an $N_2$ inert atmosphere at a heating rate of 10 °C/min from 50 °C to 800 °C.

**Supporting Information**

Analysis to estimate chemical formula of RhB@ZIF-71, SEM images, X-ray diffraction patterns and crystallinity analysis, Raman spectra, excitation-emission spectra, lifetime emission spectra, spectrofluorometer characterization techniques, physically mixed sample of RhB + ZIF-71.

**Conflicts of Interest**

There are no conflicts to declare

**Acknowledgements**

We thank the Research Complex at Harwell (RCaH) for access to advanced materials characterization facilities. J.C.T., M.G., and A.K.C. thank the ERC Consolidator Grant under the grant agreement 771575 (PROMOFS) for supporting the research.

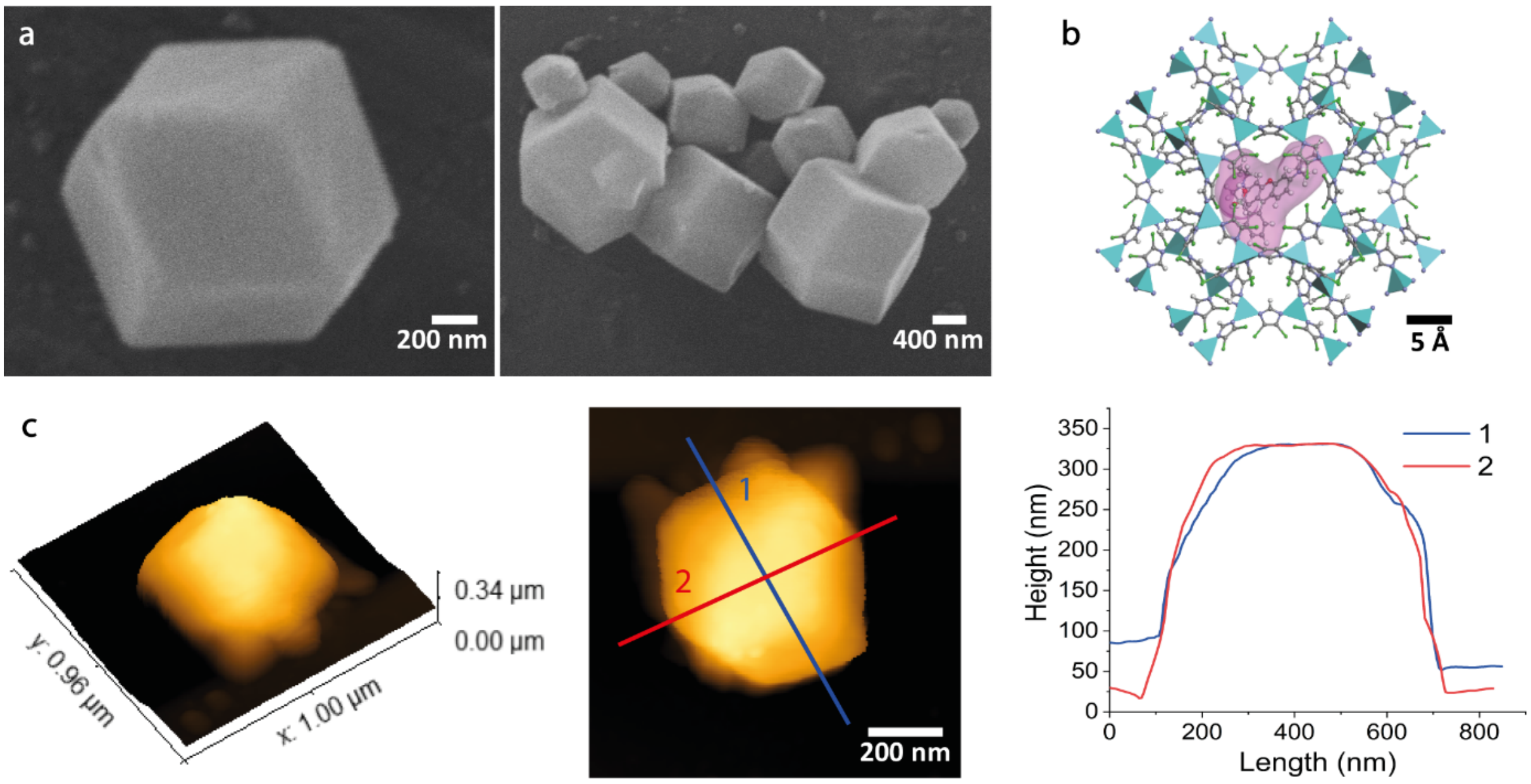

**Figure 1.** (a) SEM micrographs of the sub-micron sized crystals of RhB@ZIF-71 exhibiting a rhombic dodecahedron habit. (b) Illustration of the guest@MOF framework structure of RhB@ZIF-71, where the pore of the ZIF-71 host is used to confine the luminescent RhB guest (represented by the molecule with red surface). From the pore spatial constraint, it can be deduced that two RhB molecules could occupy a unit cell of ZIF-71. Color scheme: $ZnN_4$ tetrahedron in cyan, nitrogen in dark blue, carbon in grey, hydrogen in white, chlorine in green, and oxygen in red. (c) AFM height topography and the corresponding cross-sectional profiles of a representative RhB@ZIF-71 crystal.

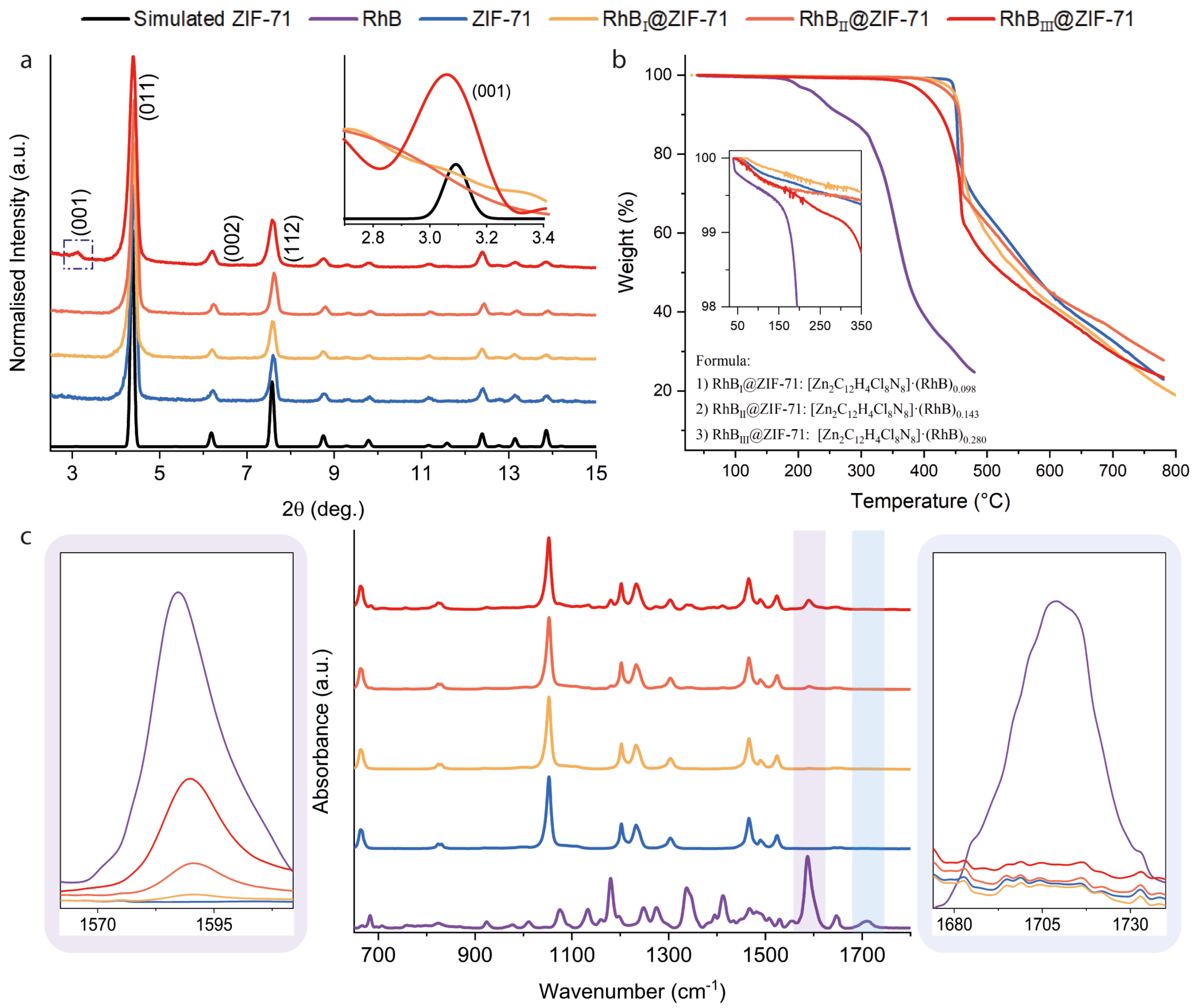

**Figure 2.** (a) Comparison of the XRD patterns of the simulated/synthesized ZIF-71 and RhB@ZIF-71 featuring three RhB concentrations. Inset: the (001) peak of simulated ZIF-71 is enlarged to compare with RhB$_{III}$@ZIF-71, while this peak is absent for both RhB$_I$@ZIF-71 and RhB$_{II}$@ZIF-71. The simulated pattern of ZIF-71 was generated from the crystallographic information file (CIF) obtained from the Cambridge Structural Database (CCDC code: GITVIP). (b) TGA and (c) ATR-FTIR results for ZIF-71, RhB, and RhB@ZIF-71. The subscripts I, II, and III designate the concentration of RhB used in the synthesis as 0.01, 0.05, and 0.5 mmol, respectively. The method to determine the chemical formulae derived from TGA data is presented in the SI section 1.

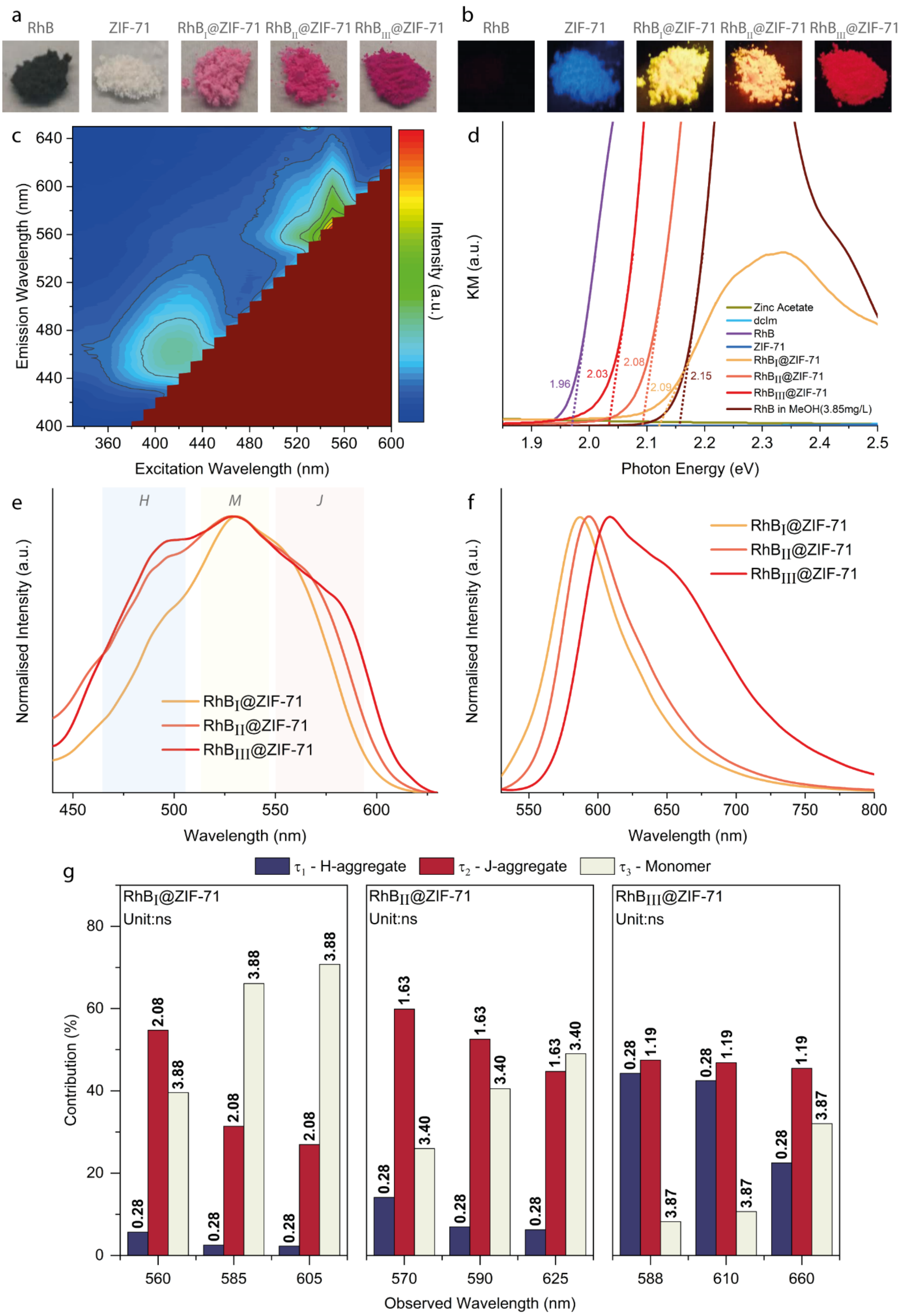

**Figure 3.** (a) RhB, ZIF-71, and RhB@ZIF-71 with different RhB concentrations seen in the visible light, and (b) their luminescence under 365 nm UV excitation. (c) Emission map of ZIF-71 powder. (d) Kubelka-Munk (KM) function for estimating the band gaps based on the photon energy intercepts. (e) Normalised excitation spectra (measured under em@650 nm), and (f) the normalised emission spectra (measured under ex@515 nm). (g) Lifetime data of RhB@ZIF-71 obtained using three different RhB concentrations.

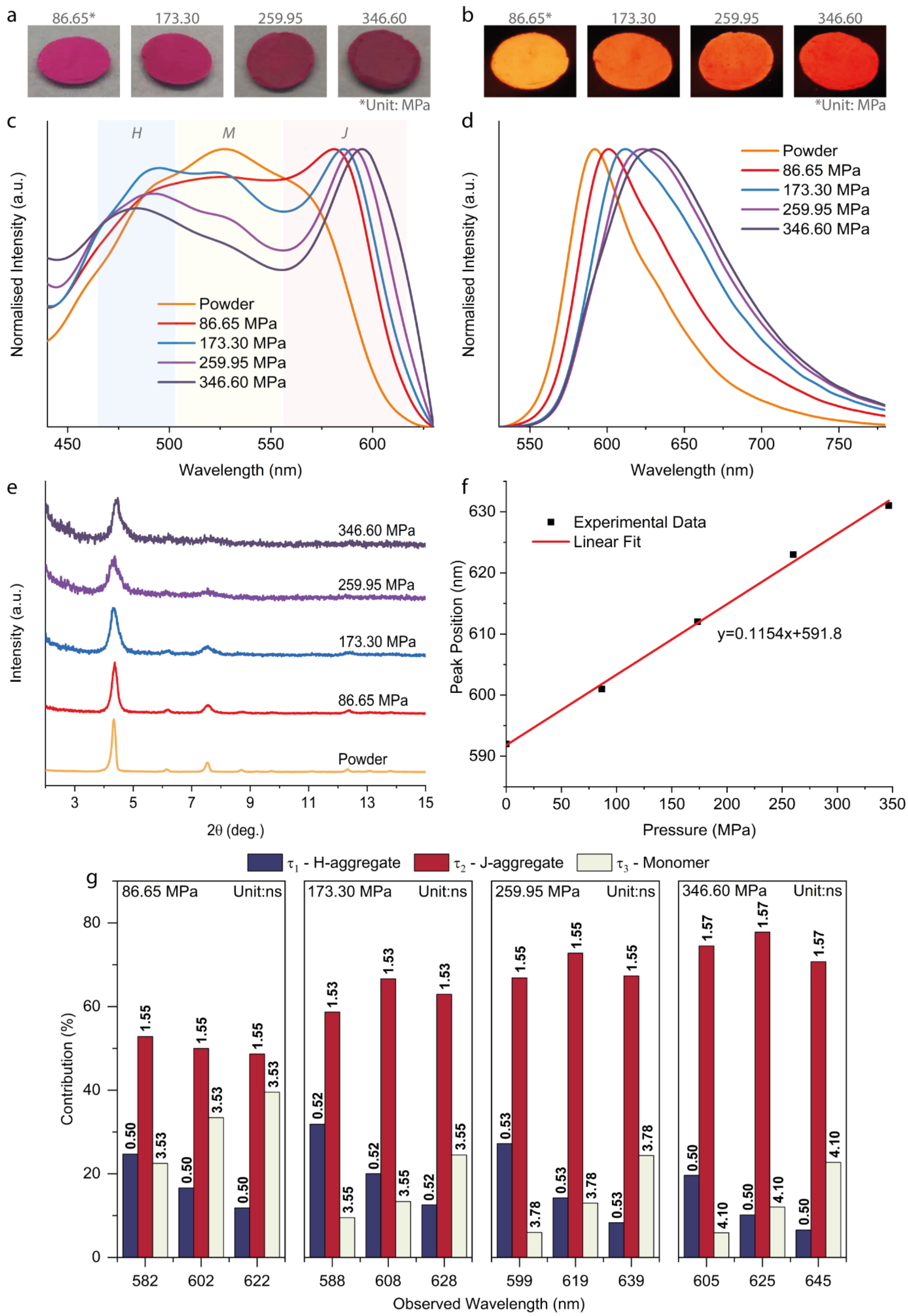

**Figure 4.** (a) $RhB_{II}$@ZIF-71 pellets prepared using different pelleting pressures, their colors viewed in visible light, and (b) their luminescence under the 365 nm UV excitation. (c) The normalised excitation spectra (measured under em@650 nm), and (d) the normalised emission spectra (measured under ex@515 nm) of the $RhB_{II}$@ZIF-71 pellets. (e) XRD of the $RhB_{II}$@ZIF-71 pellets. (f) Linear relationship between the emission peak wavelength and the applied pressure for $RhB_{II}$@ZIF-71. (g) Lifetime data of $RhB_{II}$@ZIF-71 pellets showing the contributions from the monomer, H- and J-aggregates.

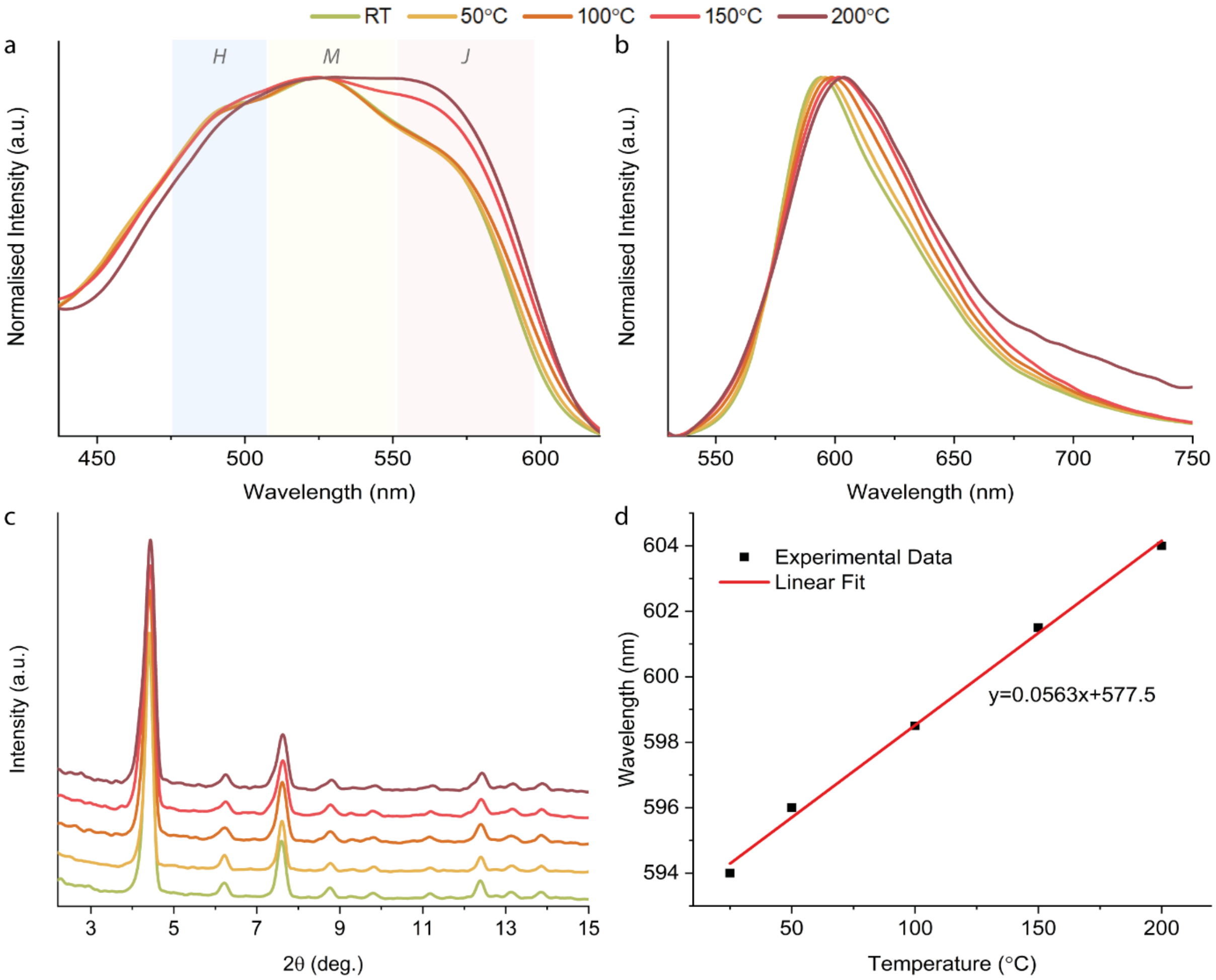

**Figure 5.** (a) Normalised excitation spectra (measured under em@650 nm) and (b) normalised emission spectra (measured under ex@365 nm) of RhB@ZIF-71 at different temperatures. (c) XRD patterns of RhB@ZIF-71 pellets after being tested at different temperatures. (d) Linear relationship of the emission peak wavelength as a function of temperature for RhB@ZIF-71.

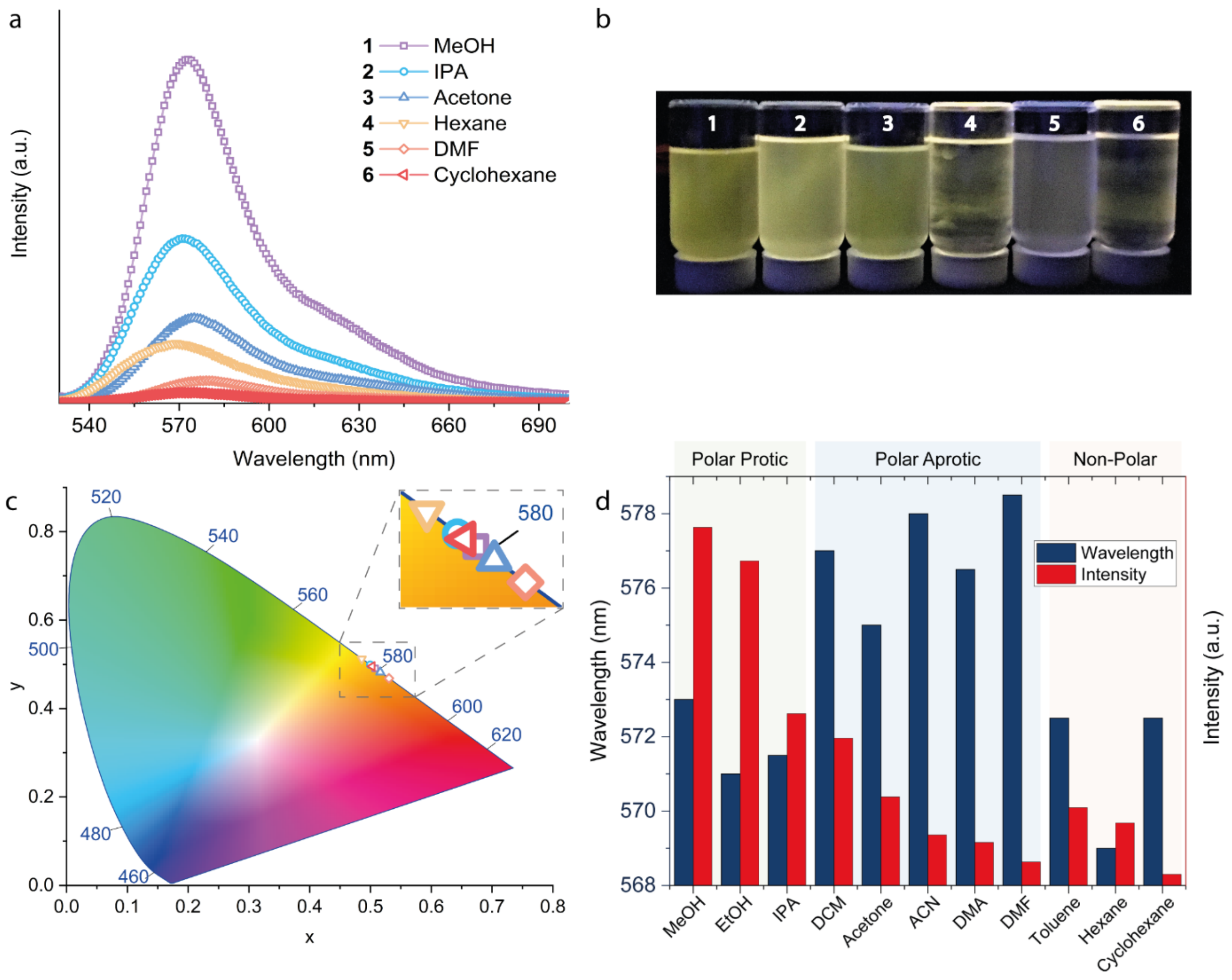

**Figure 6.** (a) Emission spectra (ex@525 nm) of $RhB_I$@ZIF-71 when exposed to different volatile organic compounds (VOCs). (b) Solvatochromism observed under the 365 nm UV excitation, where the concentration ratio used was 1 mg of $RhB_I$@ZIF-71 dispersed in 20 mL of solvent. (c) Color variation presented on the CIE 1931 chromaticity diagram. (d) Change in emission wavelength and peak intensity of RhB@ZIF-71 when tested in a wide range of polar protic, polar aprotic, and non-polar solvents.

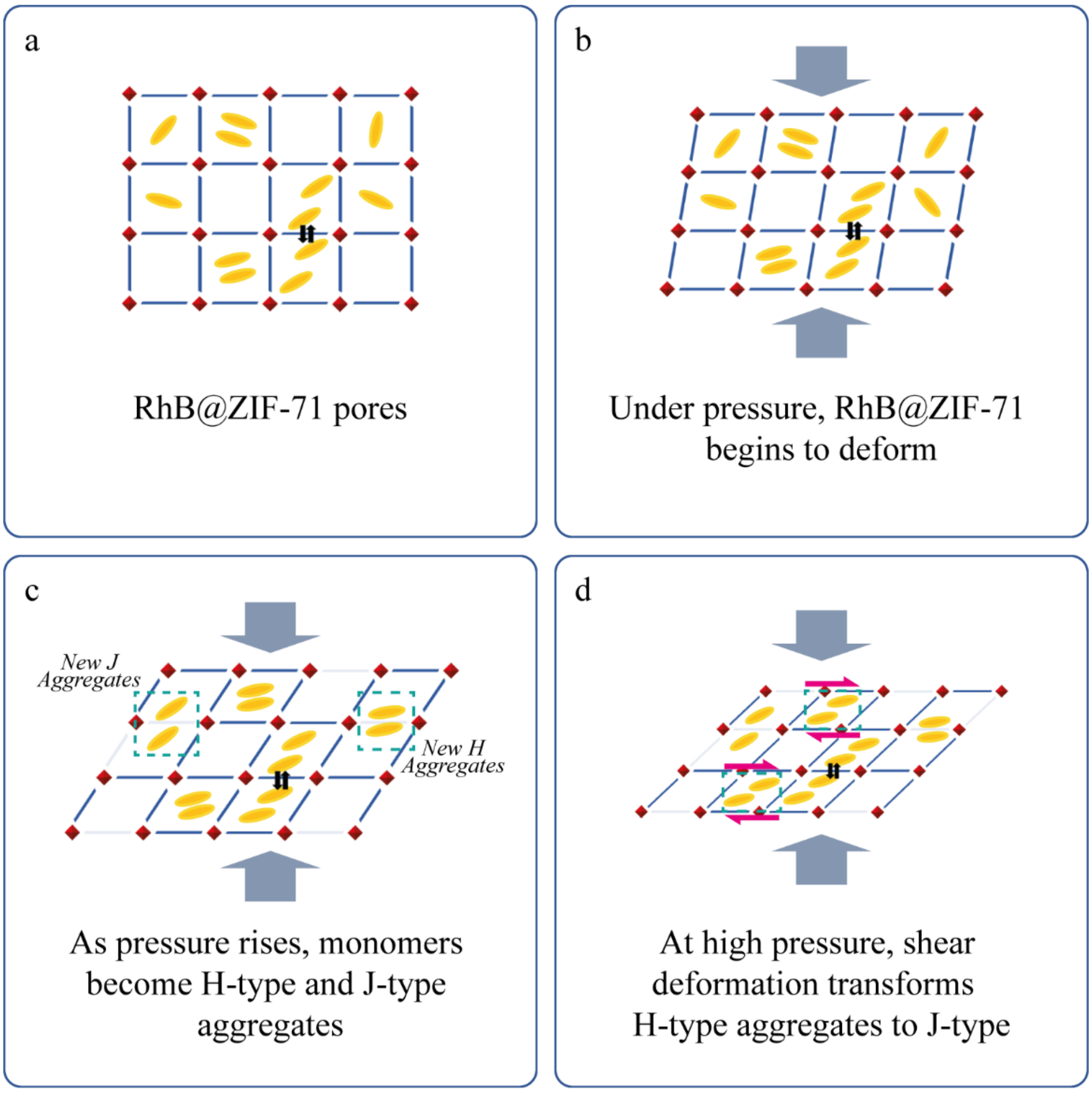

**Scheme 1.** Proposed deformation of RhB@ZIF-71 subject to a mechanical pressure, where the red nodes and blue rods represent $ZnN_4$ coordination environment and dcIm linker, respectively. Yellow ellipses represent the RhB guests within the pore of ZIF-71 host. (a) The possible interactions between J-aggregates are denoted by the pair of black arrows. (b) and (c) illustrate how the structural distortion under pressure cause the formation of new H- and J-type aggregates. The grey rods represent the broken linkers under pressure. (d) Due to shear deformation (denoted by antiparallel pink arrows), the majority of H-type aggregates transform to J-type.

**Table 1.** Values of time constants ($\tau_i$), normalised pre-exponential factors ($a_i$), and fractional contributions ($c_i = \tau_i \cdot a_i$) of the emission decay of pristine dcIm linker in solid state and in methanol solutions upon excitation at 362.5 nm ($R_t = \Sigma a_i e^{(-t/\tau_i)}$, $R_t$ is the quantity/counts at time *t*).

| dcIm | $\lambda_{obs}$ [nm] | $\tau_1$ [ns] | $a_1$ | $c_1$ [%] | $\tau_2$ [ns] | $a_2$ | $c_2$ [%] | $\tau_3$ [ns] | $a_3$ | $c_3$ [%] | $\chi^2$ |
|---|---|---|---|---|---|---|---|---|---|---|---|
| Solid state | 470 | 0.36 | 0.041 | 11.89 | 1.76 | 0.038 | 53.95 | 4.44 | 0.010 | 34.16 | 1.052 |
| 0.0365 M | 442 | 0.36 | 0.057 | 17.57 | 1.76 | 0.033 | 49.49 | 4.44 | 0.009 | 32.94 | 1.115 |
| 0.5 M | 423 | 0.36 | 0.061 | 20.28 | 1.76 | 0.034 | 55.09 | 4.44 | 0.006 | 24.63 | 1.258 |
| | 443 | 0.36 | 0.060 | 18.80 | 1.76 | 0.037 | 55.64 | 4.44 | 0.007 | 25.55 | 1.082 |
| | 463 | 0.36 | 0.058 | 18.26 | 1.76 | 0.036 | 54.73 | 4.44 | 0.007 | 27.02 | 1.168 |

**Table 2.** Values of time constants ($\tau_i$), normalised pre-exponential factors ($a_i$), and fractional contributions ($c_i = \tau_i \cdot a_i$) of the emission decay of pristine ZIF-71 powder upon excitation at 362.5 nm.

| ZIF-71 | $\lambda_{obs}$ [nm] | $\tau_1$ [ns] | $a_1$ | $c_1$ [%] | $\tau_2$ [ns] | $a_2$ | $c_2$ [%] | $\tau_3$ [ns] | $a_3$ | $c_3$ [%] | $\chi^2$ |
|---|---|---|---|---|---|---|---|---|---|---|---|
| Solid state | 450 | 0.58 | 0.049 | 26.43 | 2.11 | 0.029 | 57.24 | 5.61 | 0.003 | 16.32 | 1.150 |
| | 558 | 0.58 | 0.041 | 14.74 | 2.11 | 0.026 | 34.50 | 5.61 | 0.015 | 50.76 | 1.215 |

**Table 3.** The quantum yield of RhB@ZIF-71 with different RhB concentrations

| Sample | QY[a)] [%] | | QY[b)] [%] | |
|---|---|---|---|---|
| | Ex@485 nm | Ex@525 nm | Ex@485 nm | Ex@525 nm |
| $RhB_I$@ZIF-71 | 23.99 | 28.25 | 35.43 | 39.53 |
| $RhB_{II}$@ZIF-71 | 13.74 | 14.15 | 17.51 | 18.30 |
| $RhB_{III}$@ZIF-71 | 1.68 | 1.85 | 2.32 | 3.31 |

[a)]Samples were directly measured; [b)]Sample (10 wt.%) were firstly mixed with $BaSO_4$ (90 wt.%) and then measured.

*Supporting Information*

*for*

# Dye-Encapsulated Zeolitic Imidazolate Framework (ZIF-71) for Fluorochromic Sensing of Pressure, Temperature, and Volatile Solvents

*Yang Zhang, Mario Gutiérrez, Abhijeet K. Chaudhari and Jin-Chong Tan**

Multifunctional Materials & Composites (MMC) Laboratory,
Department of Engineering Science, University of Oxford,
Parks Road, Oxford OX1 3PJ, United Kingdom.

*Corresponding author:

jin-chong.tan@eng.ox.ac.uk

## Characterization Techniques

### 1. Calculation of the chemical formula of RhB@ZIF-71 and the number of RhB molecules per pore by means of thermogravimetric analysis (TGA) data

To calculate the chemical formula of the RhB@ZIF-71 materials, we firstly estimated the mass percentage of ZIF-71 and RhB corresponding to their mass loss in the TGA curves (Fig. 2b in main manuscript). In the case of $RhB_I$@ZIF-71, the loss of mass associated to ZIF-71 and RhB was determined to be 93.5% and 7.5%, respectively. For $RhB_{II}$@ZIF-71 and $RhB_{III}$@ZIF-71 composites, the mass loss of ZIF-71 was 90.8% and 83.4%, respectively, while for RhB was 9.2% and 16.6%, respectively. From those values, we have calculated the number of moles of RhB per moles of ZIF-71 by applying the following equations:

$$n = \frac{m}{M} \qquad (1)$$

where *n* stands for the number of moles, *m* is the mass, and *M* is the molecular weight. From the equation (1) it is possible to estimate the ratio between the moles of ZIF-71 and RhB:

$$\frac{n_{ZIF-71}}{n_{RhB}} = \frac{\frac{m_{ZIF-71}}{M_{ZIF-71}}}{\frac{m_{RhB}}{M_{RhB}}} = \frac{m_{ZIF-71}}{m_{RhB}} \times \frac{M_{RhB}}{M_{ZIF-71}} \qquad (2)$$

Knowing that the chemical formula of ZIF-71 is $[Zn_2C_{12}H_4Cl_8N_8]$, the global chemical formula of RhB@ZIF-71 is given by:

(I) $RhB_I$@ZIF-71:

$$\frac{n_{ZIF-71}}{n_{RhB}} = \frac{93.5}{7.5} \times \frac{479}{674} = \frac{1}{0.098}$$

**$[Zn_2C_{12}H_4Cl_8N_8]\cdot(RhB)_{0.098}$**

(II) $RhB_{II}$@ZIF-71:

$$\frac{n_{ZIF-71}}{n_{RhB}} = \frac{90.8}{9.2} \times \frac{479}{674} = \frac{1}{0.143}$$

**$[Zn_2C_{12}H_4Cl_8N_8]\cdot(RhB)_{0.143}$**

(III) $RhB_{III}$@ZIF-71:

$$\frac{n_{ZIF-71}}{n_{RhB}} = \frac{83.4}{16.6} \times \frac{479}{674} = \frac{1}{0.280}$$

**$[Zn_2C_{12}H_4Cl_8N_8]\cdot(RhB)_{0.280}$**

Once calculated the amount or RhB per two atoms of Zn (i.e. $Zn_2$), and knowing that a single pore of ZIF-71 contains 24 atoms of Zn, we can estimate the mean number of RhB molecules per pore in each material:

(I) $RhB_{I}$@ZIF-71 → 0.098 × 12 = **1.2 RhB molecules per pore**

(II) $RhB_{II}$@ZIF-71 → 0.143 × 12 = **1.7 RhB molecules per pore**

(III) $RhB_{III}$@ZIF-71 → 0.280 × 12 = **3.4 RhB molecules per pore**

Because of the spatial constraint of the ZIF-71 pore, only a maximum two RhB guest molecules could occupy a pore volume. On this basis, we reasoned that (I) and (II) are confined inside the pores of ZIF-71, whereas (III) has excess molecules that are adhering to the outer surface of the ZIF-71 crystals.

## 2. Fluorescence spectroscopy using the Edinburgh Instruments FS-5 setup

i. Steady-state excitation (solid) in module SC-10: for the measurement of excitation spectra, the emission wavelength was selected at 650 nm. The dwell time was 0.2 s, the step size was 1 nm and 2 scans were performed for each measurement.

ii. Steady state emission (solid) in module SC-10: for the measurement of emission spectra, the excitation wavelength for RhB@ZIF-71 was 515 nm. The dwell time was 0.2 s, the step size was 1 nm and 2 scans were performed for each measurement.

iii. Lifetime TCSPC measurement (solid): A 362.5 nm laser was used for lifetime measurement. The stop condition was set to be at 10,000 counts.

iv. QY measurement (solid) in module SC-30: the starting scan wavelength was selected as 20 nm before the excitation wavelength. QY analysis was done using the 'Fluoracle' software.

v. Steady-state thermochromism measurement (solid) in module SC-28: The luminescent properties of the sample were measured when the set temperature was reached for 10 minutes (the instrumental settings were the same as 1 and 2 above).

vi. Mechanochromism measurement (solid): The RhB@ZIF-71 powder was pressed into pellets using the Specac hydraulic press with pellet Ø1 cm. Then the excitation/emission spectra and lifetime of these pellets were measured (the instrumental settings were the same as 1 and 2 above).

vii. Characterization of solutions in module SC-05 or SC-20: Liquid solutions were performed in quartz cuvettes. For excitation, emission, lifetime and QY measurements, all the instrumental settings were the same as solid-state measurements described above.

viii. Solvatochromism measurement: After the RhB@ZIF-71 powder was combined with a solvent, it was sonicated for 10 minutes to obtain a good dispersion. The luminescent properties of the solution were measured using instrumental settings as in 1 and 2 above.

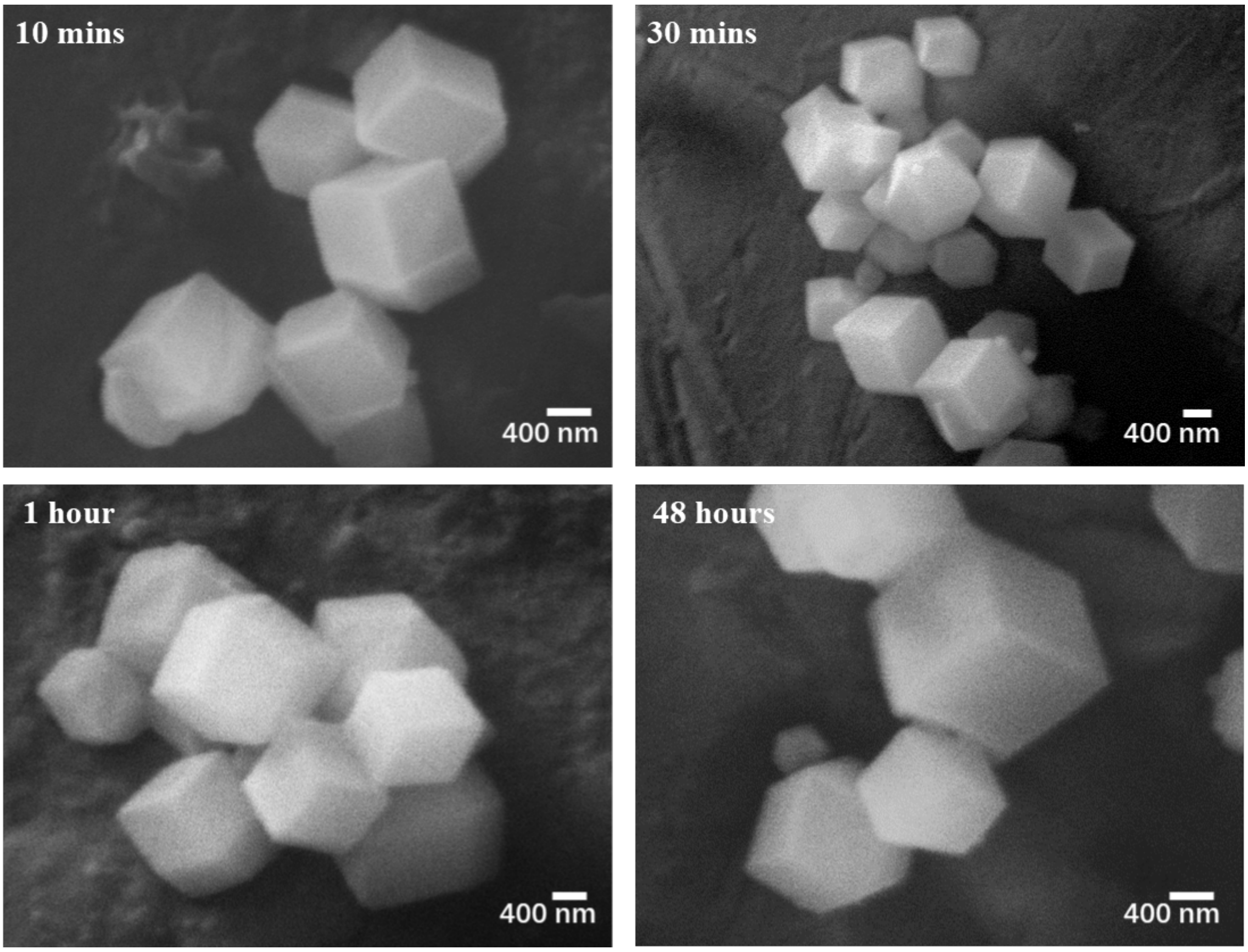

**Figure S1.** The SEM images of RhB@ZIF-71 prepared using different reaction times, ranging from 10 mins to 48 hours.

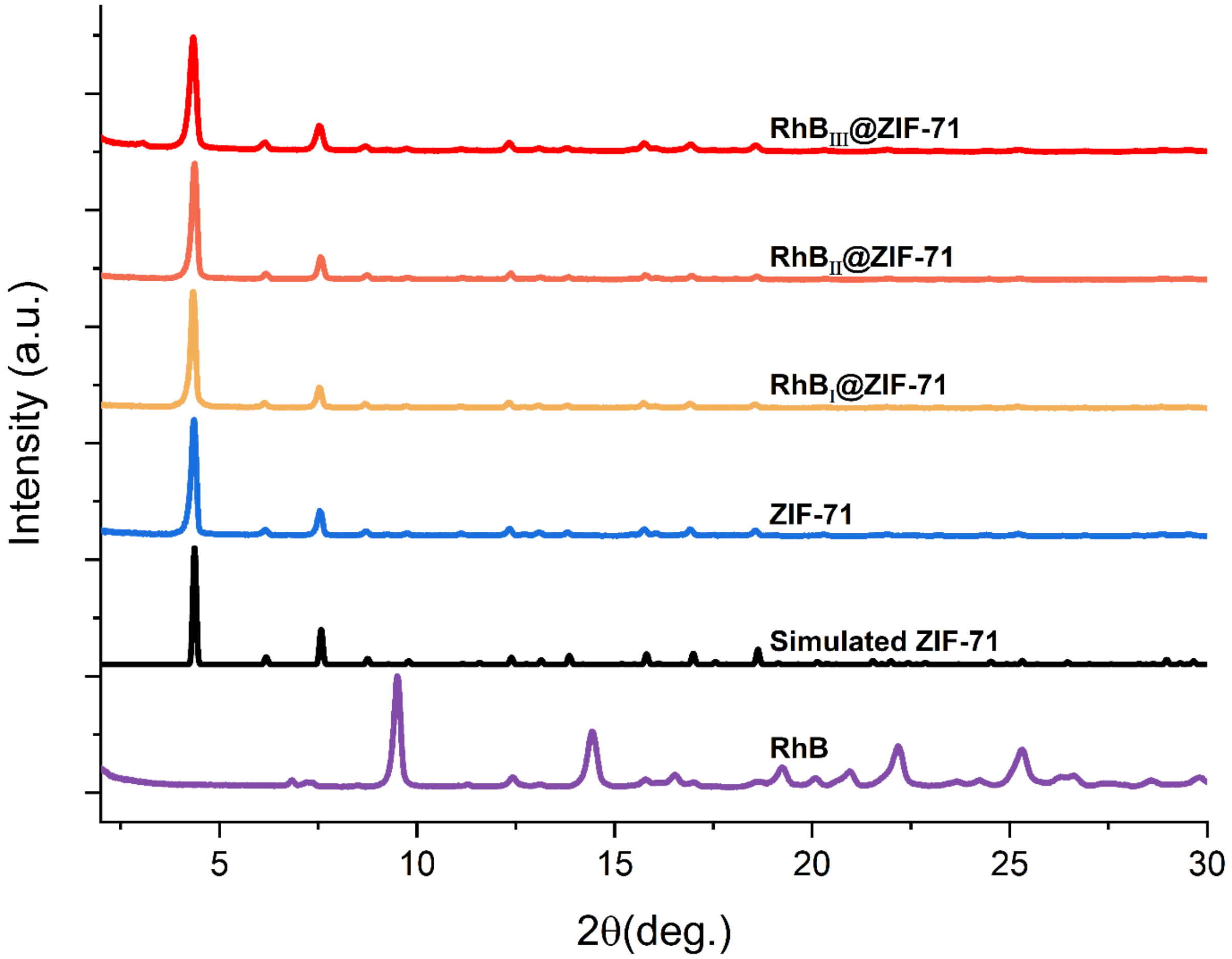

**Figure S2.** XRD patterns of pure RhB, pristine ZIF-71 crystals, and RhB@ZIF-71 guest-host composites prepared using three different concentrations of RhB guests. The subscripts I, II, and III correspond to the three concentrations of RhB used in the synthesis: 0.01, 0.05, and 0.5 mmol, respectively. The simulated pattern of ZIF-71 was generated from the crystallographic information file (CIF) obtained from the Cambridge Structural Database (CCDC code: GITVIP).

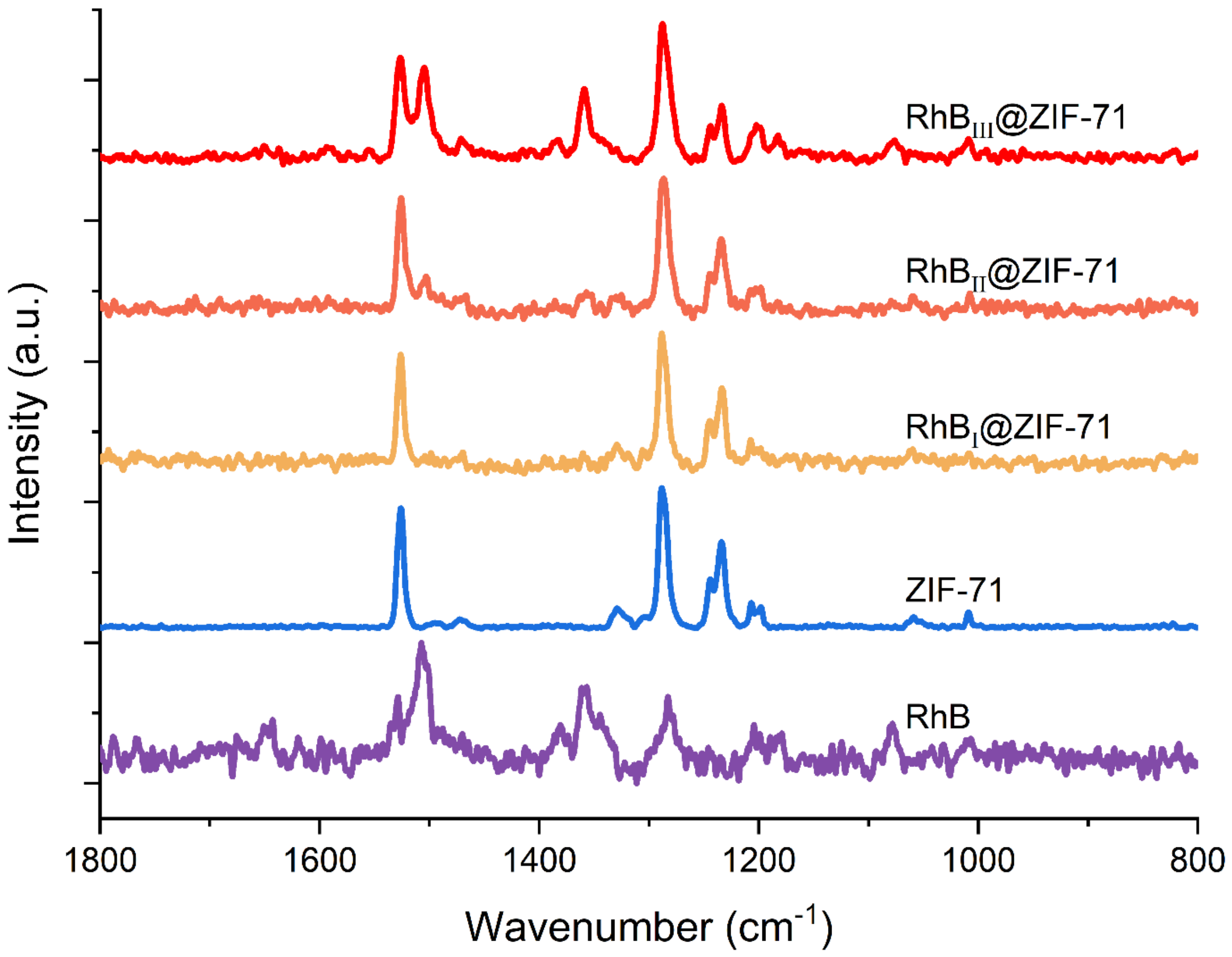

**Figure S3.** Raman spectra of pure RhB, pristine ZIF-71 crystals, and RhB@ZIF-71 guest-host composites prepared using three different concentrations of RhB guests.

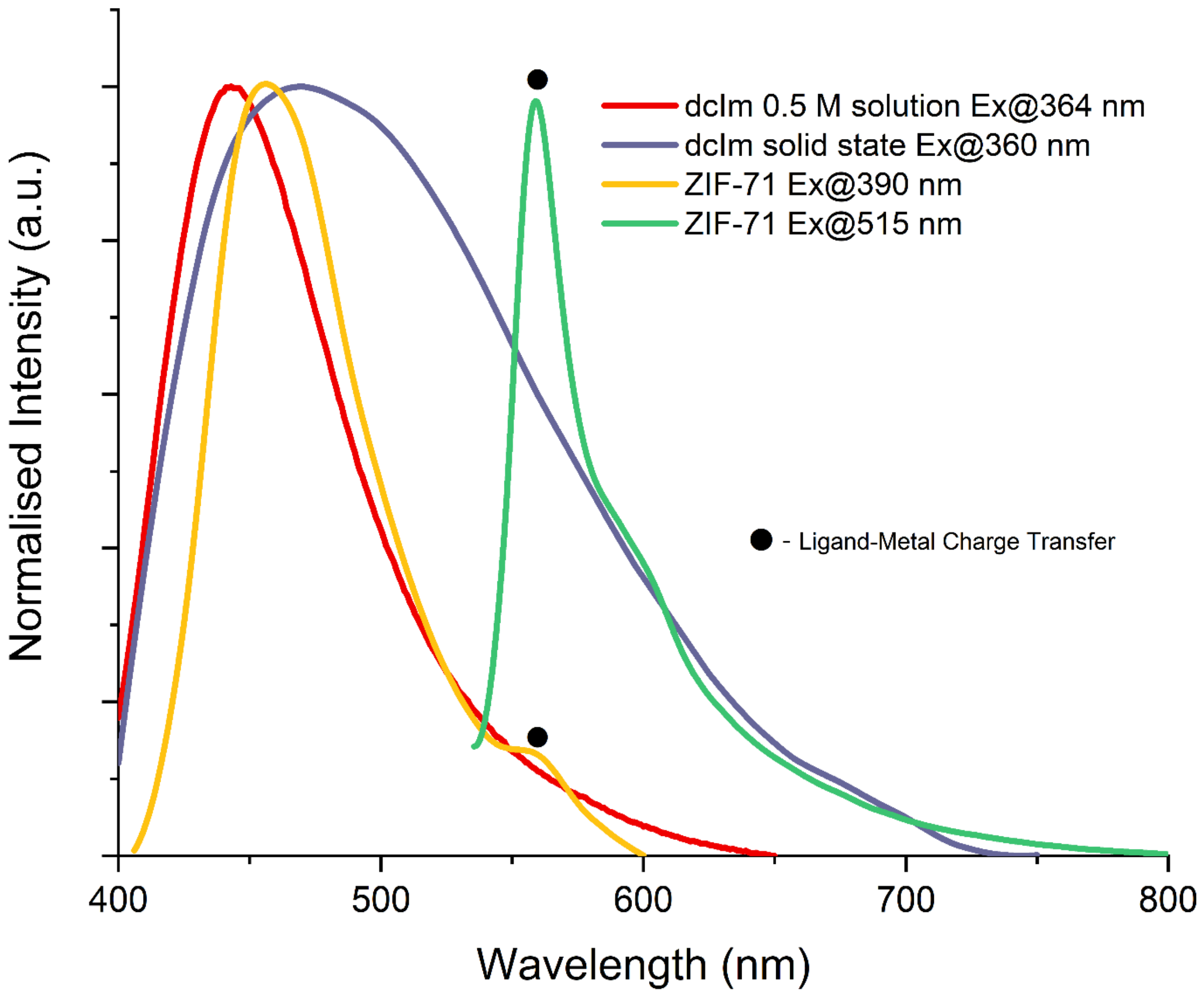

**Figure S4.** Normalised emission spectra of the dcIm linker in methanol solution (0.5 M) and in the solid state, compared to that of pristine ZIF-71 in solid state excited at 390 nm or 515 nm.

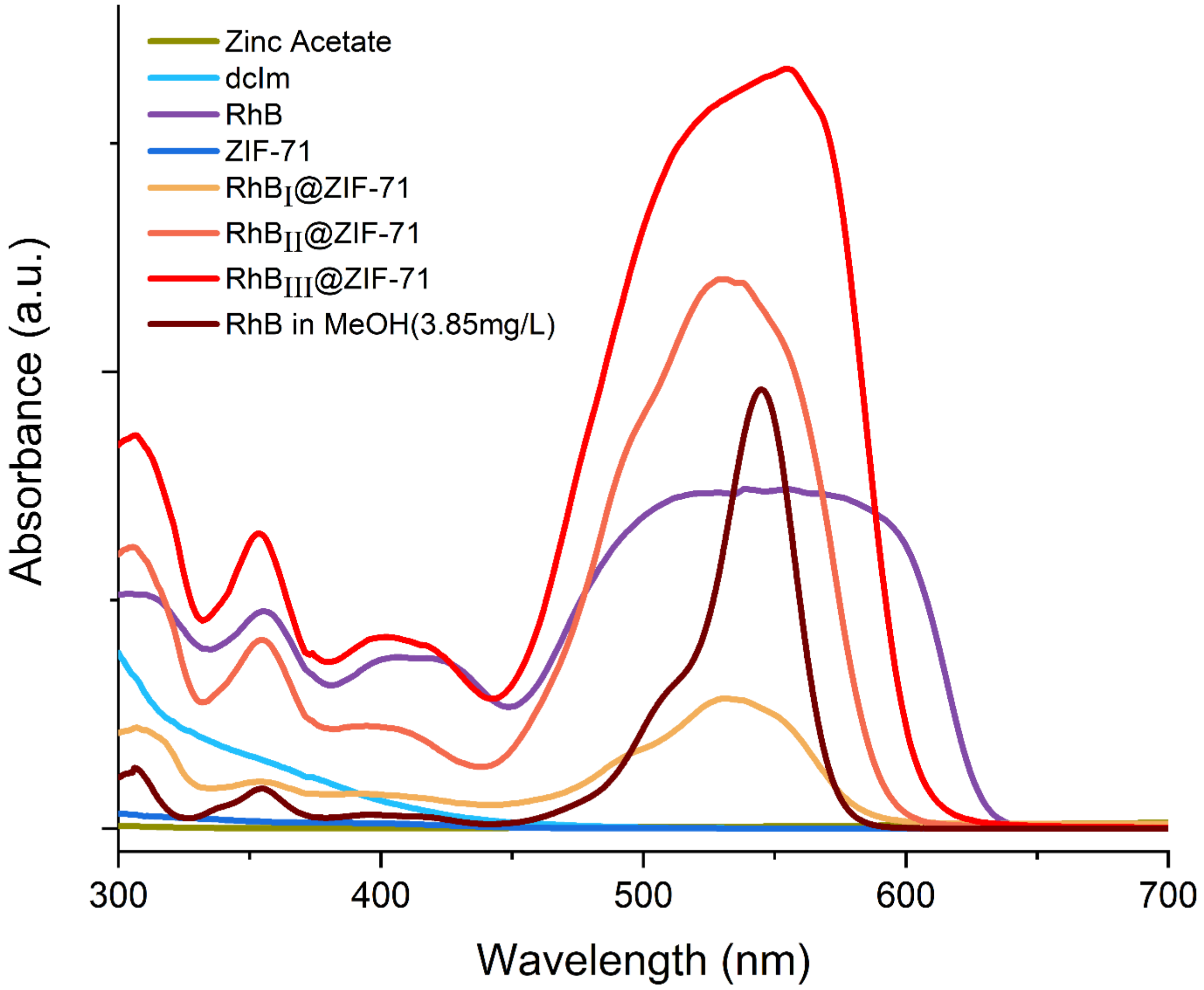

**Figure S5.** Diffuse reflectance spectra (DRS) of the absorption bands

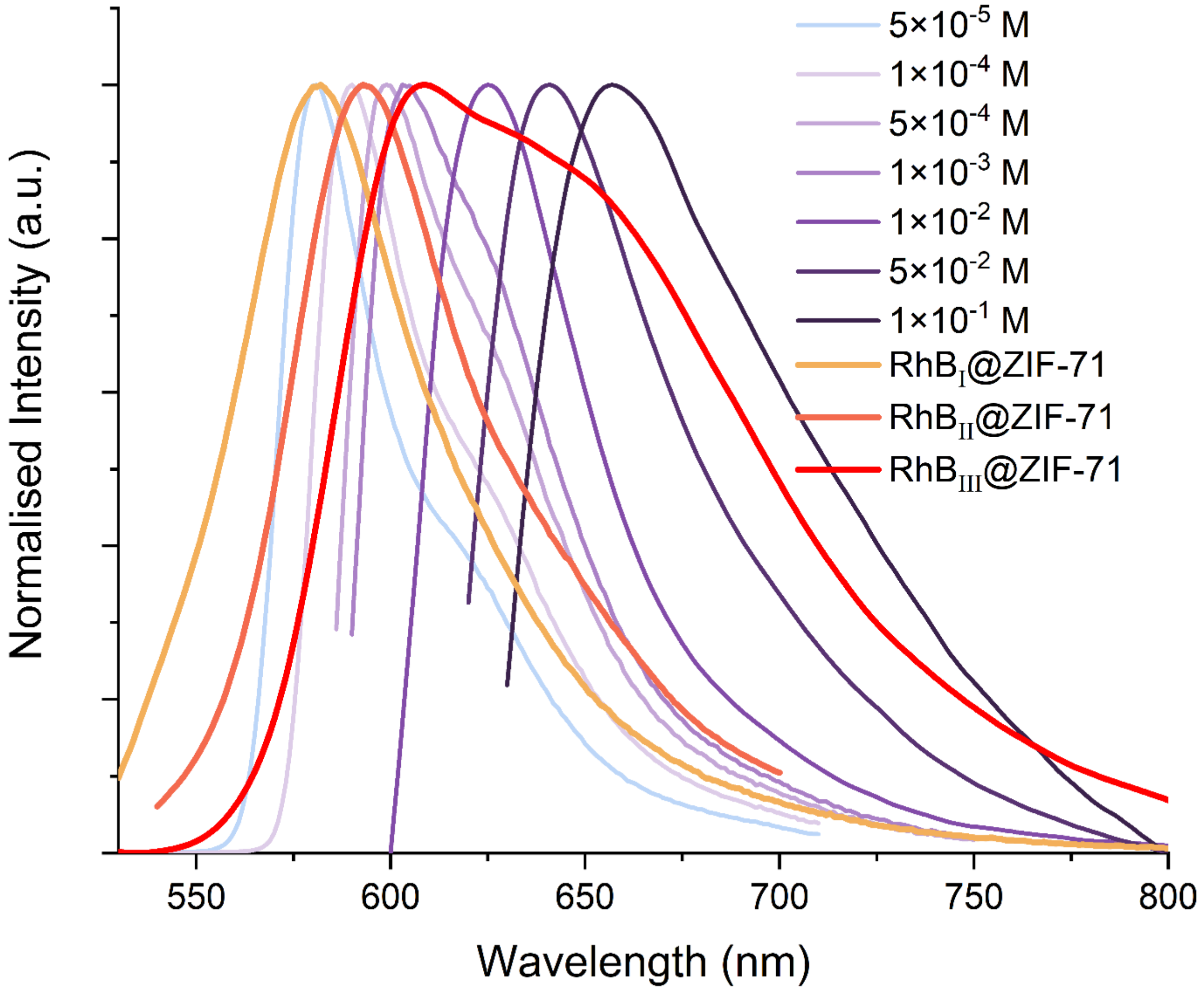

**Figure S6.** Normalised emission spectra of RhB in MeOH solution (ex@515 nm) of different concentrations, and RhB$_{I/II/III}$@ZIF-71 powders in the solid state.

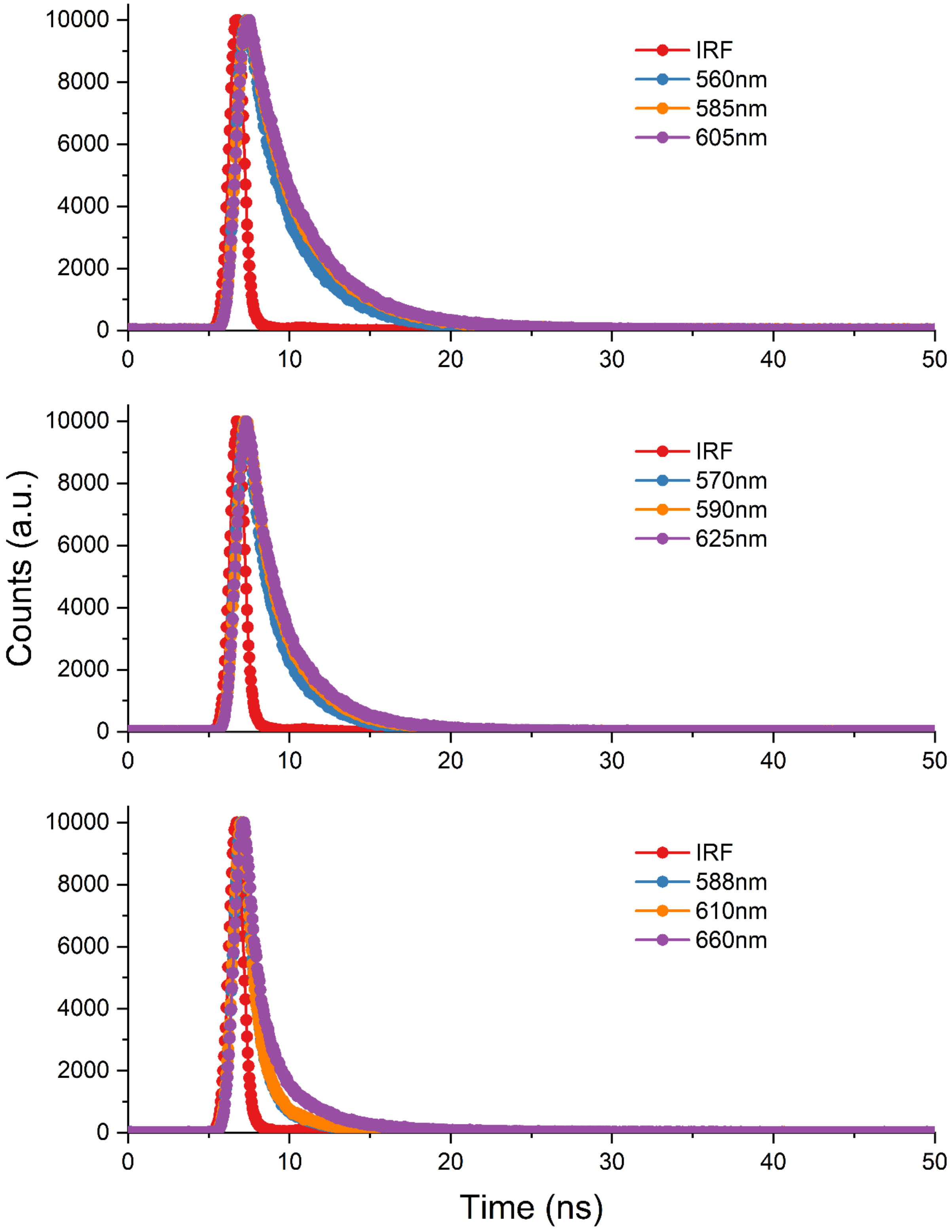

**Figure S7.** Lifetime emission spectra showing the decays of (a) $RhB_I$@ZIF-71, (b) $RhB_{II}$@ZIF-71, (c) $RhB_{III}$@ZIF-71.

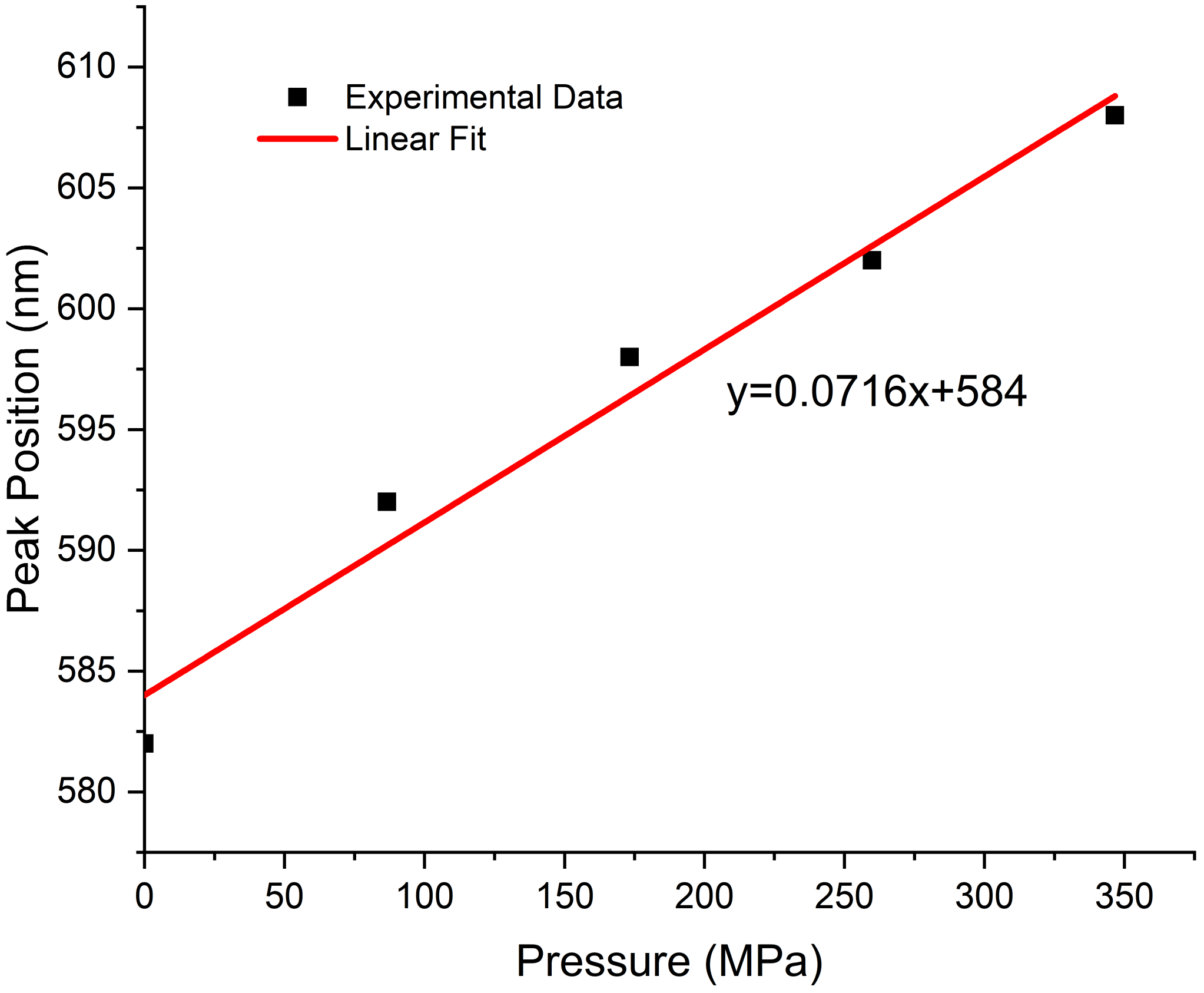

**Figure S8.** Correlation between the emission peak wavelength and the applied pressure for preparing the $RhB_I$@ZIF-71 pellets.

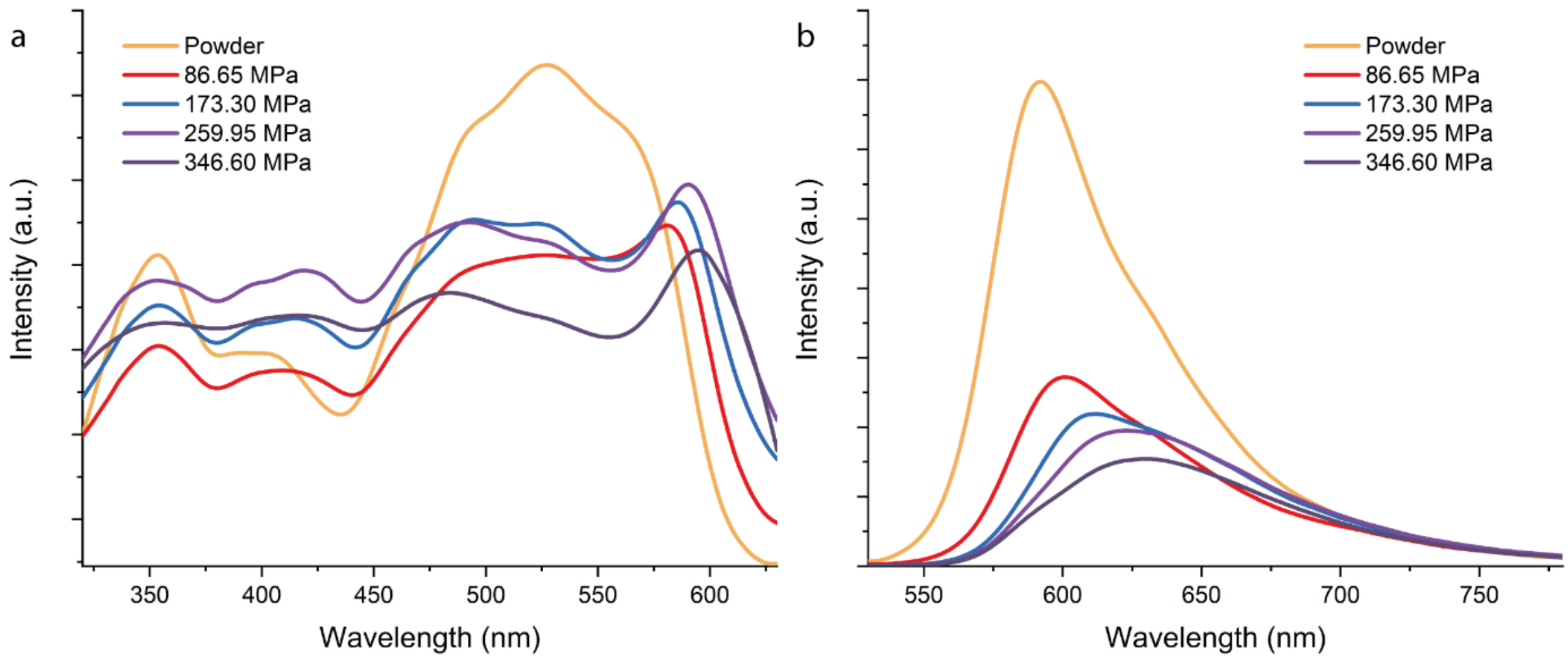

**Figure S9.** (a) Excitation spectra (measured under em@650 nm), and (b) emission spectra (measured under ex@515 nm) of $RhB_{II}$@ZIF-71 pellets.

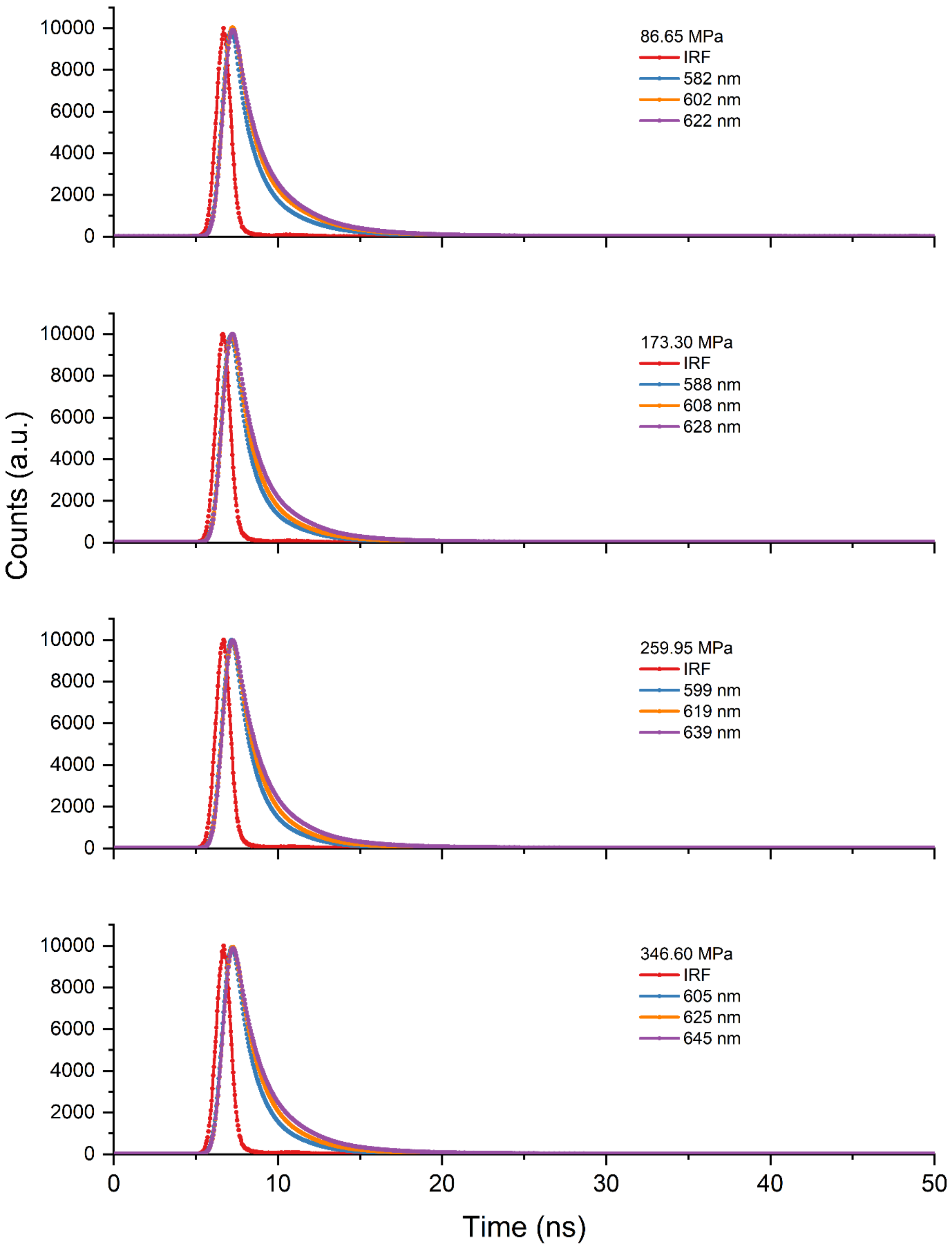

**Figure S10.** Emission decays of pellets under (a) 86.65 MPa, (b) 173.30 MPa, (c) 259.95 MPa, and (d) 346.60 MPa.

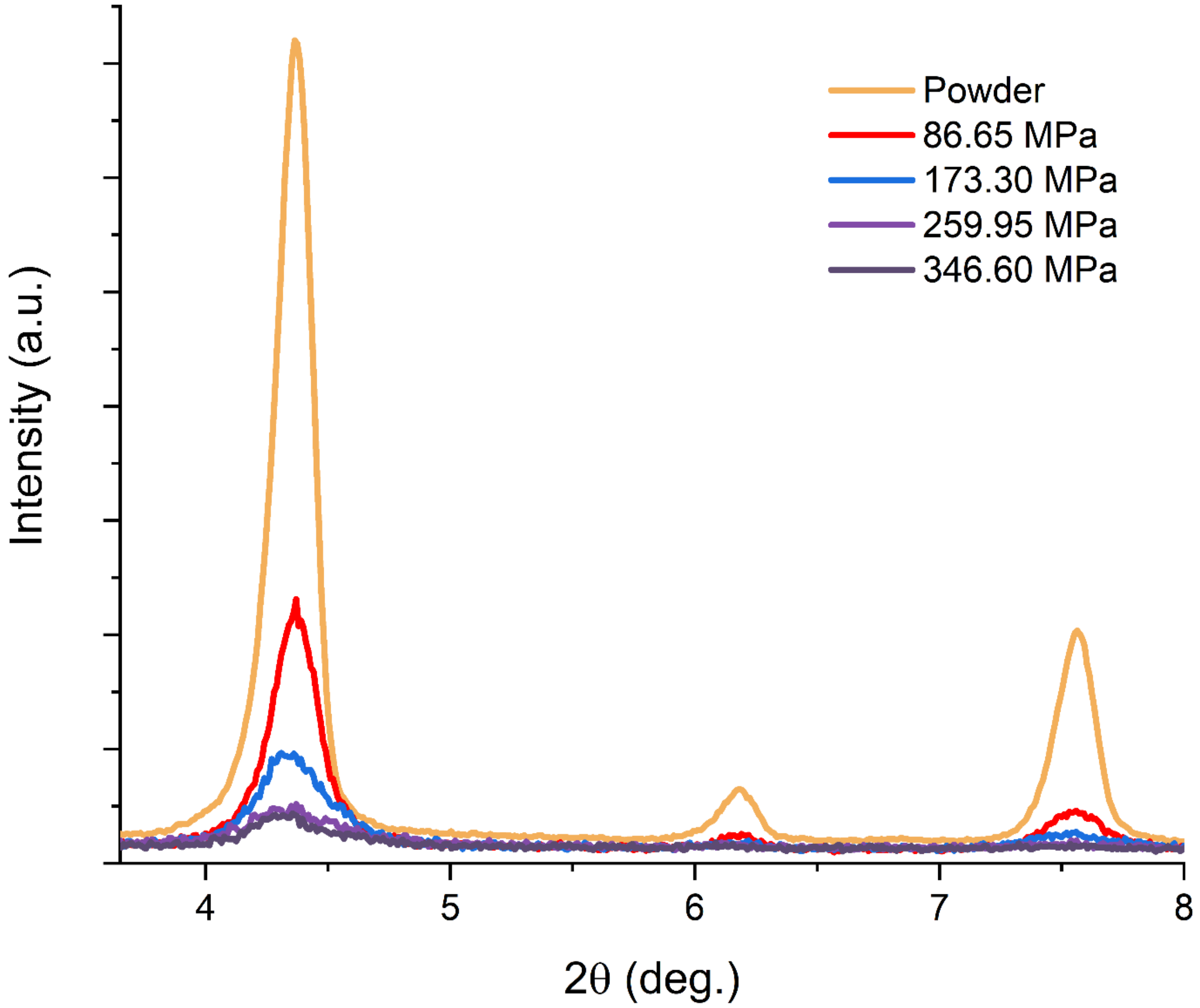

**Figure S11.** The unnormalized XRD patterns of the $RhB_{II}$@ZIF-71 pellets.

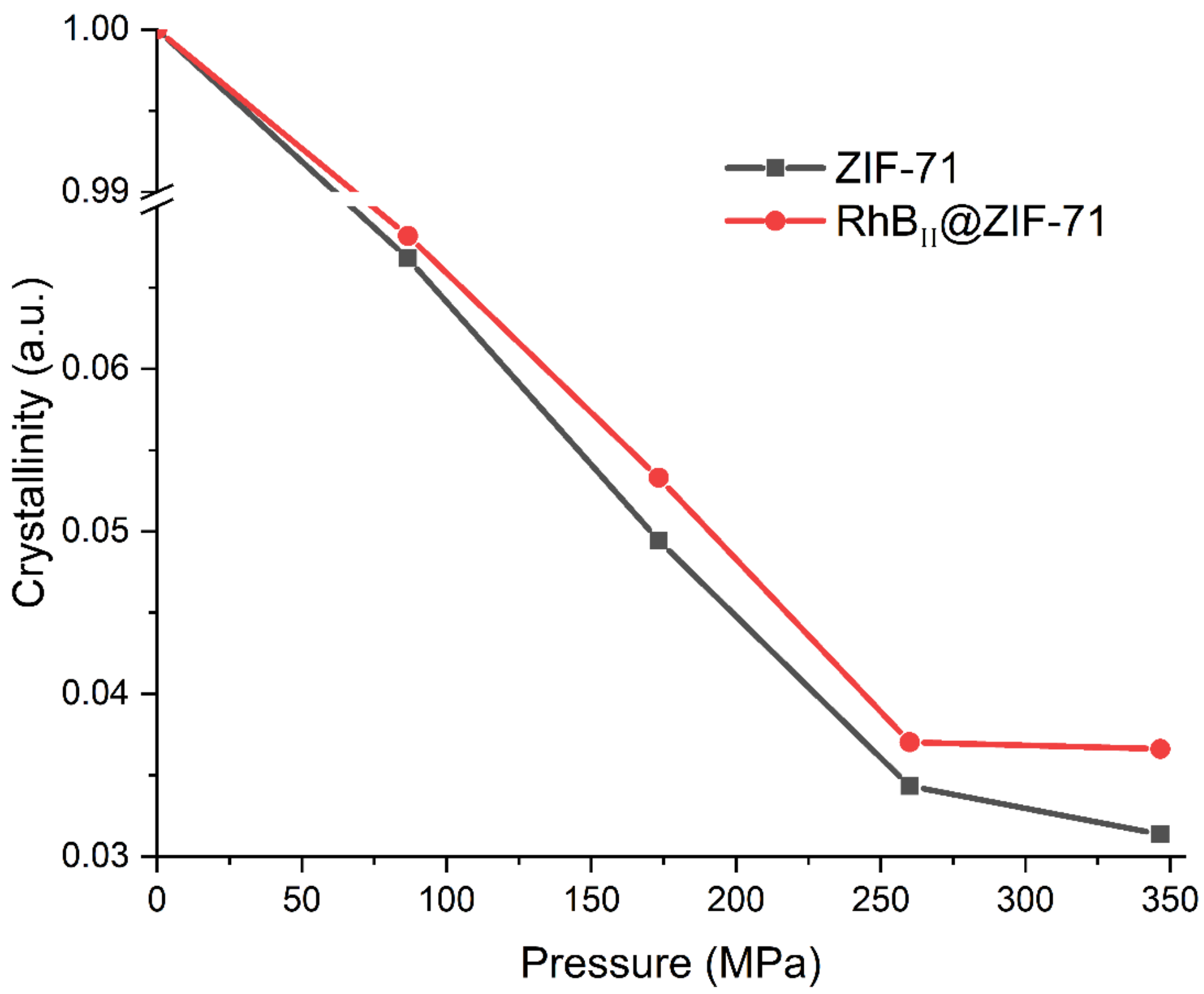

**Figure S12**. Crystallinity of ZIF-71 and RhB$_{II}$@ZIF-71 as a function of pressure, where data are presented in Table S4.

According to the Scherrer equation, the size of the crystalline domains $D$ can be calculated by:

$$D = \frac{K\,\lambda}{\Delta\cos\theta}$$

where $K$ is a constant, $\lambda$ is the wavelength, $\Delta$ is the FWHM, and $\theta$ is the diffraction angle of the corresponding diffraction peak.

Therefore, the crystallinity can be obtained by using $D_{\mathrm{p}}/D_{\mathrm{initial}}$ [1], where $D_{\mathrm{p}}$ is the size of the crystalline domains after being subjected to pressure $P$, and $D_{\mathrm{initial}}$ is the domain size of the pristine powder.

$$\therefore \text{crystallinity} = \frac{D_{\mathrm{P}}}{D_{\mathrm{initial}}} = \frac{\dfrac{K\,\lambda}{\Delta_{\mathrm{P}}\;\cos\theta}}{\dfrac{K\,\lambda}{\Delta_{\mathrm{initial}}\cos\theta}} = \frac{\Delta_{\mathrm{initial}}}{\Delta_{\mathrm{P}}}$$

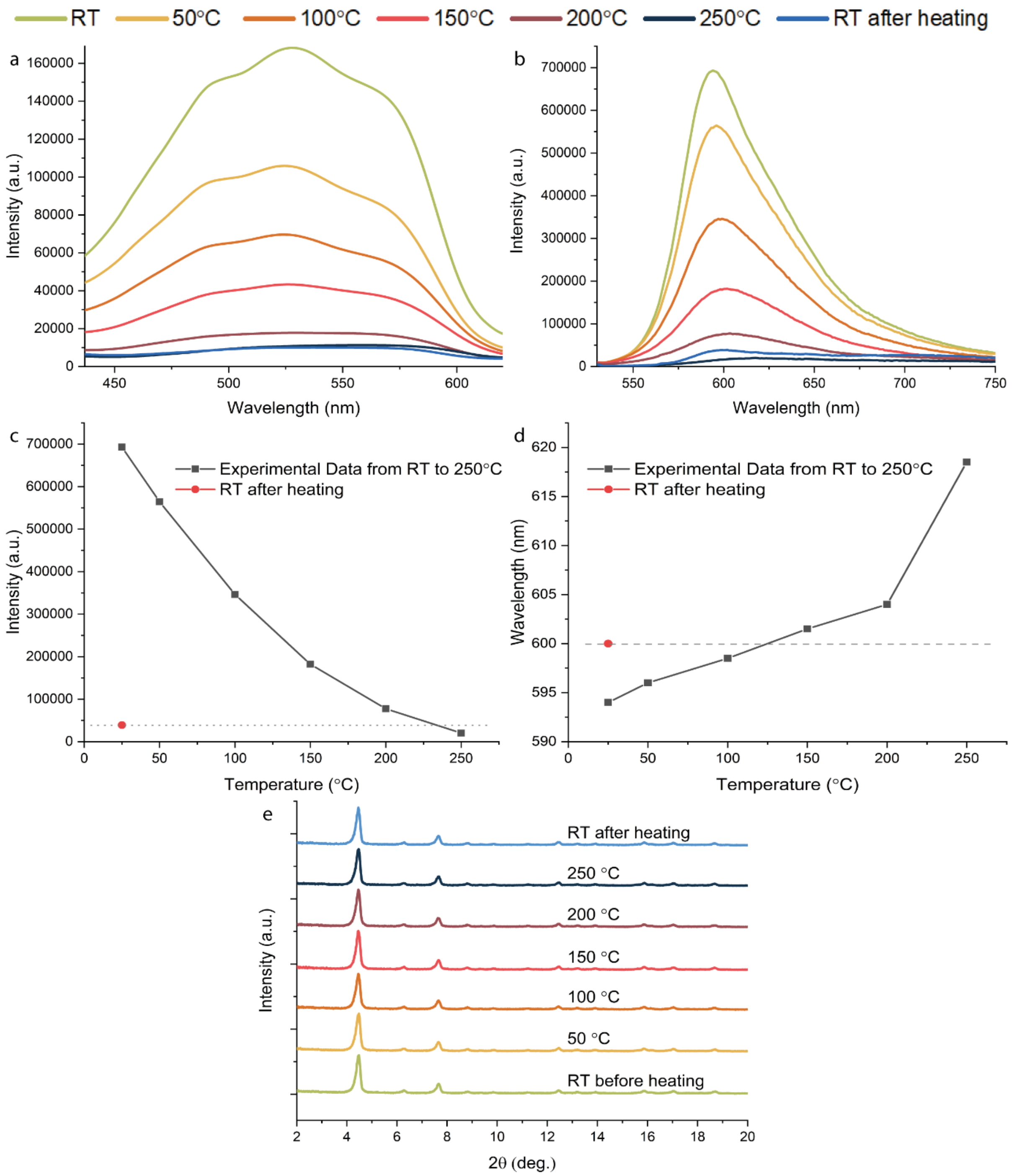

**Figure S13.** (a) Excitation spectra (measured under em@650 nm) and (b) emission spectra (measured under ex@365 nm) of $RhB_{II}$@ZIF-71 determined at different temperatures. The variation of (c) peak intensity and (d) peak wavelength as a function of temperature. (e) XRD patterns of $RhB_{II}$@ZIF-71 at different temperatures.

| Peak Wavelength [nm] | Solvents | | | | | | | | | | |
|---|---|---|---|---|---|---|---|---|---|---|---|
| | MeOH | EtOH | IPA | DCM | Acetone | Toluene | Hexane | ACN | DMA | DMF | Cyclohexane |
| RhB$_I$@ZIF-71 | 573 | 571 | 571.5 | 577 | 575 | 572.5 | 569 | 578 | 576.5 | 578.5 | 572.5 |
| RhB | 573 | 570 | 564 | 577 | /[a)] | /[b)] | /[b)] | 564.5 | /[a)] | /[a)] | /[b)] |

[a)]The signal was too low to be detected. [b)]RhB is poorly soluble or insoluble

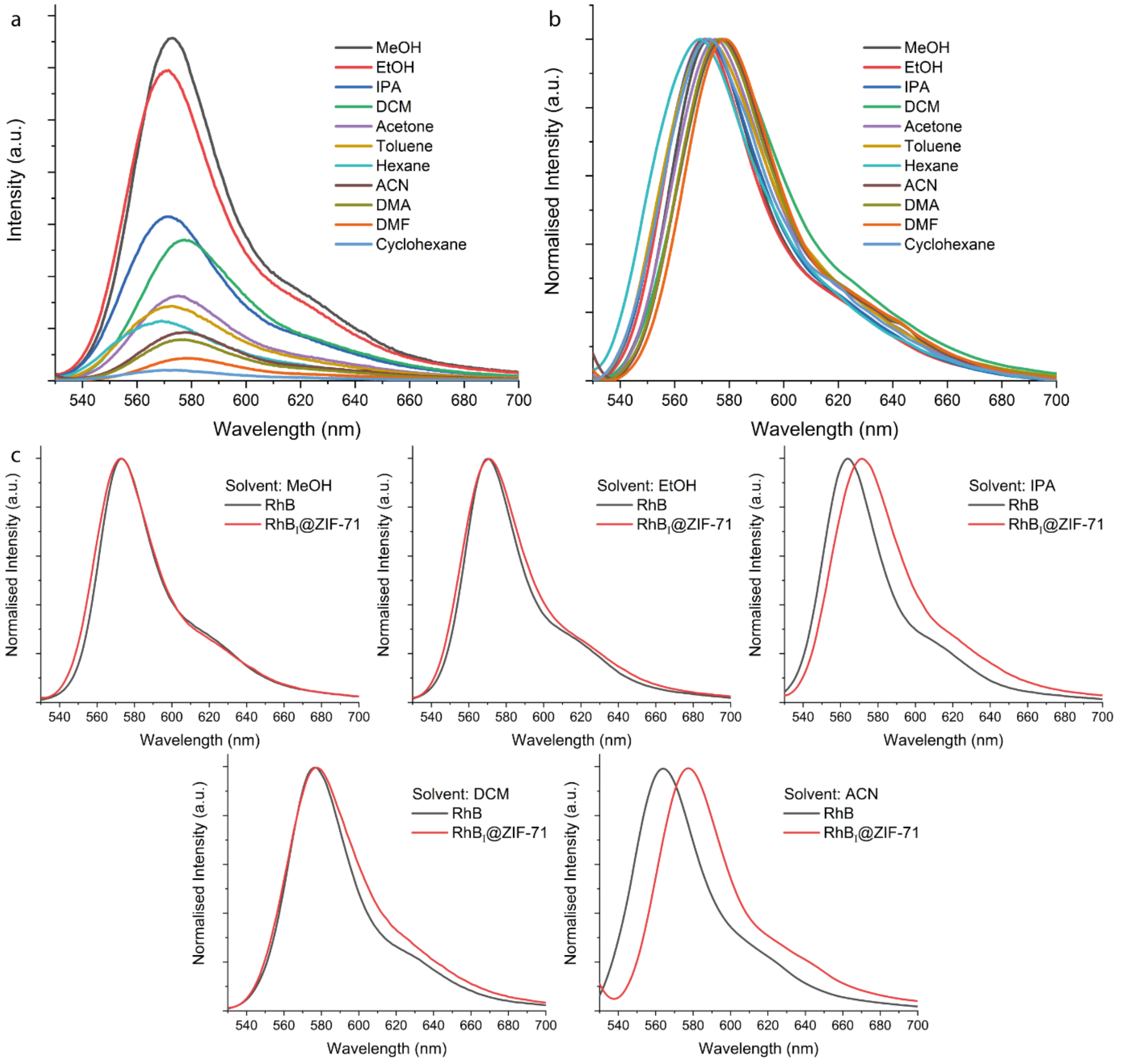

**Figure S14.** (a) Emission spectra (unnormalized) and (b) normalised emission spectra of RhB$_I$@ZIF-71 in different solvents (each comprising 1 mg of RhB$_I$@ZIF-71 in 20 ml solvent; ex@525 nm). (c) Comparison of the emission of RhB$_I$@ZIF-71 solutions and pure RhB in different solvents ($7.5 \times 10^{-6}$ M, ex@525 nm).

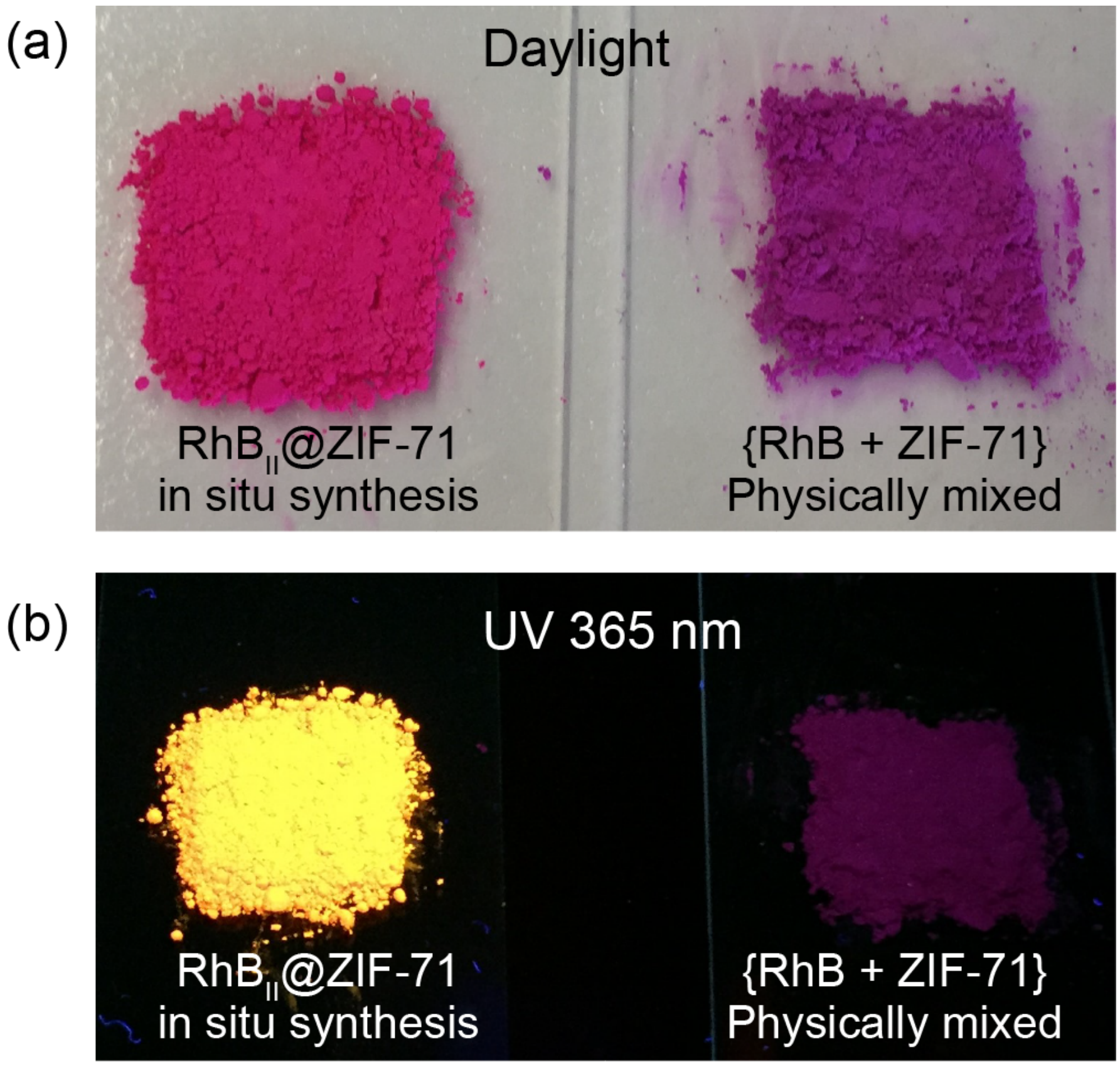

**Figure S15.** Comparing the sample colour and luminescence behaviour observed (a) under daylight and (b) in UV irradiation. Marked differences can be seen between the samples of RhB$_{II}$@ZIF-71 derived from *in situ* encapsulation synthesis (see method in manuscript), in contrast to the {RhB + ZIF-71} sample made by simple *physical mixing* or blending of the two constituents (in mortar and pestle by maintaining the same guest concentration as RhB$_{II}$@ZIF-71). The physically mixed 'composite' is not photoluminescent under UV irradiation due to quenching of the RhB guests aggregating on the outer surface of the ZIF-71 crystals. Conversely, it is clear that the RhB guest molecules confined in the ZIF-71 pores are well isolated and thus highly luminescent under UV excitation.

**Table S1.** The quantum yield (QY) of ZIF-71 in the solid state.

| Sample | QY [%] | | |
|---|---|---|---|
| | Ex@385 nm | Ex@485 nm | Ex@525 nm |
| ZIF-71 | 9.13 | Undetectable[a)] | Undetectable[a)] |

[a)]Too weak to be detected

**Table S2.** Values of time constants ($\tau_i$), normalised pre-exponential factors ($a_i$), and fractional contributions ($c_i = \tau_i \cdot a_i$) of the emission decay of RhB@ZIF-71 with different concentration of RhB in solid state and in methanol solutions upon excitation at 362.5 nm.

| Sample | $\lambda_{obs}$ [nm] | $\tau_1$ [ns] | $a_1$ | $c_1$ [%] | $\tau_2$ [ns] | $a_2$ | $c_2$ [%] | $\tau_3$ [ns] | $a_3$ | $c_3$ [%] | $\chi^2$ |
|---|---|---|---|---|---|---|---|---|---|---|---|
| $RhB_I$@ZIF-71 | 560 | 0.28 | 0.030 | 5.67 | 2.08 | 0.039 | 54.75 | 3.88 | 0.015 | 39.58 | 1.098 |
| | 585 | 0.28 | 0.016 | 2.50 | 2.08 | 0.027 | 31.41 | 3.88 | 0.030 | 66.09 | 1.132 |
| | 605 | 0.28 | 0.015 | 2.27 | 2.08 | 0.024 | 26.97 | 3.88 | 0.034 | 70.76 | 1.242 |
| $RhB_{II}$@ZIF-71 | 570 | 0.28 | 0.060 | 14.13 | 1.63 | 0.043 | 59.88 | 3.40 | 0.009 | 25.99 | 1.164 |
| | 590 | 0.28 | 0.033 | 6.96 | 1.63 | 0.042 | 52.54 | 3.40 | 0.015 | 40.50 | 1.114 |
| | 625 | 0.28 | 0.032 | 6.24 | 1.63 | 0.038 | 44.73 | 3.40 | 0.020 | 49.03 | 1.120 |
| $RhB_{III}$@ZIF-71 | 588 | 0.28 | 0.124 | 44.27 | 1.19 | 0.031 | 47.49 | 3.87 | 0.002 | 8.24 | 1.292 |
| | 610 | 0.28 | 0.121 | 42.50 | 1.19 | 0.031 | 46.84 | 3.87 | 0.002 | 10.66 | 1.141 |
| | 660 | 0.28 | 0.082 | 22.50 | 1.19 | 0.039 | 45.47 | 3.87 | 0.008 | 32.04 | 1.154 |

**Table S3.** Values of time constants ($\tau_i$), normalised pre-exponential factors ($a_i$), and fractional contributions ($c_i = \tau_i \cdot a_i$) of the emission decay of RhB$_{II}$@ZIF-71 pellets upon excitation at 362.5 nm.

| Pressure [MPa] | $\lambda_{obs}$ [nm] | $\tau_1$ [ns] | $a_1$ | $c_1$ [%] | $\tau_2$ [ns] | $a_2$ | $c_2$ [%] | $\tau_3$ [ns] | $a_3$ | $c_3$ [%] | $\chi^2$ |
|---|---|---|---|---|---|---|---|---|---|---|---|
| 86.65 | 582 | 0.50 | 0.053 | 24.70 | 1.55 | 0.036 | 52.82 | 3.53 | 0.007 | 22.48 | 1.119 |
| | 602 | 0.50 | 0.040 | 16.60 | 1.55 | 0.038 | 49.98 | 3.53 | 0.011 | 33.42 | 1.113 |
| | 622 | 0.50 | 0.030 | 11.81 | 1.55 | 0.040 | 48.66 | 3.53 | 0.014 | 39.52 | 1.081 |
| 173.30 | 588 | 0.52 | 0.059 | 31.82 | 1.53 | 0.037 | 58.67 | 3.55 | 0.003 | 9.51 | 1.186 |
| | 608 | 0.52 | 0.040 | 20.02 | 1.53 | 0.046 | 66.63 | 3.55 | 0.004 | 13.35 | 1.104 |
| | 628 | 0.52 | 0.028 | 12.57 | 1.53 | 0.048 | 62.93 | 3.55 | 0.008 | 24.50 | 1.094 |
| 259.95 | 599 | 0.53 | 0.052 | 27.20 | 1.55 | 0.043 | 66.84 | 3.78 | 0.002 | 5.96 | 1.082 |
| | 619 | 0.53 | 0.029 | 14.25 | 1.55 | 0.051 | 72.76 | 3.78 | 0.004 | 12.99 | 1.085 |
| | 639 | 0.53 | 0.019 | 8.32 | 1.55 | 0.053 | 67.34 | 3.78 | 0.008 | 24.34 | 1.043 |
| 346.60 | 605 | 0.50 | 0.040 | 19.61 | 1.57 | 0.048 | 74.49 | 4.10 | 0.001 | 5.90 | 1.234 |
| | 625 | 0.50 | 0.023 | 10.17 | 1.57 | 0.057 | 77.78 | 4.10 | 0.003 | 12.05 | 1.146 |
| | 645 | 0.50 | 0.016 | 6.55 | 1.57 | 0.054 | 70.72 | 4.10 | 0.007 | 22.73 | 1.211 |

**Table S4.** Comparison of FWHM between pure ZIF-71 and RhB@ZIF-71.

| Sample | Pressure [MPa] | | | | |
|---|---|---|---|---|---|
| | 0 | 86.65 | 173.30 | 259.95 | 346.60 |
| ZIF-71 | 0.01610 ±0.00023 | 0.2410 ±0.0023 | 0.3257 ±0.0029 | 0.4690 ±0.0026 | 0.5130 ±0.0232 |
| $RhB_{II}$@ZIF-71 | 0.01625 ±0.0003 | 0.2383 ±0.0023 | 0.3048 ±0.0051 | 0.4387 ±0.0101 | 0.4437 ±0.0589 |